%% file: hhkm-main.tex
\documentclass[12pt,a4paper,fleqn]{article}
\usepackage[utf8]{inputenc}
\usepackage{authblk}
\usepackage{geometry}
\usepackage{xcolor}
\usepackage{natbib}
\usepackage{graphicx}
\usepackage{pdflscape}
\usepackage{appendix}
\usepackage{setspace}
\usepackage{colortbl}
\providecommand{\cellclbench}{\cellcolor[rgb]{0.9, 0.9, 0.9}}
\providecommand{\cellcl}{\cellcolor[rgb]{0.98, 0.6, 0.6}}

\usepackage{threeparttable}
\usepackage{booktabs}
\usepackage{enumitem}

\usepackage{amsmath}
\usepackage{amsfonts}
\usepackage{bm}

\usepackage{dcolumn}
\newcolumntype{d}[1]{D{.}{.}{#1}}

\makeatletter
\renewenvironment{thebibliography}[1]
         {\section*{\refname}%
          \@mkboth{\MakeUppercase\refname}{\MakeUppercase\refname}%
          \list{\@biblabel{\@arabic\c@enumiv}}%
               {\settowidth\labelwidth{\@biblabel{#1}}%
                \leftmargin\labelwidth
                \advance\leftmargin\labelsep
                \@openbib@code
                \usecounter{enumiv}%
                \let\p@enumiv\@empty
                \itemsep=0pt
                \parsep=0pt
                \leftmargin=\parindent
                \itemindent=-\parindent
                \renewcommand\theenumiv{\@arabic\c@enumiv}}%
          \sloppy
          \clubpenalty4000
          \@clubpenalty \clubpenalty
          \widowpenalty4000%
          \sfcode`\.\@m}
         {\def\@noitemerr
           {\@latex@warning{Empty `thebibliography' environment}}%
          \endlist}
\makeatother

\usepackage{textcomp}

\usepackage{caption}
\usepackage{subcaption}
\definecolor{nblue}{HTML}{000660}
\usepackage[colorlinks=true,urlcolor=nblue,linkcolor=nblue,citecolor=nblue]{hyperref}

\title{\Large{\textbf{Bayesian Modeling of TVP-VARs \\ Using Regression Trees}\thanks{
\noindent Corresponding author: Niko Hauzenberger. Department of Economics, University of Salzburg. \textit{Address}: M\"{o}nchsberg 2a, 5020 Salzburg, Austria. \textit{Email}: \href{mailto:niko.hauzenberger@plus.ac.at}{niko.hauzenberger@plus.ac.at}. This paper was previously circulated as \textit{``Bayesian Modeling of Time-Varying Parameters Using Regression Trees."} We would like to thank Todd Clark, Domenico Giannone, Ed Knotek, Dimitris Korobilis, Eoghan O'Neill, Helga Wagner, and participants at the Econometrics seminar series at the University of Glasgow, the $16^{th}$ International Conference Computational and Financial Econometrics (CFE 2022), and the Bayes@Austria 2020 workshop for comments and suggestions. The first two authors gratefully acknowledge financial support from the Austrian Science Fund (FWF, grant no. ZK 35) and the Oesterreichische Nationalbank (OeNB, Anniversary Fund, project no. 18304, 18763, and 18765). The views expressed herein are those of the authors and not necessarily those of the Federal Reserve Bank of Cleveland or the Federal Reserve System.}}}
\author{\large{
\uppercase{Niko Hauzenberger,$^{1,2}$} 
\uppercase{Florian Huber,$^1$} \\\vspace*{-0.5em}
\uppercase{Gary Koop},$^2$ and \uppercase{James Mitchell}$^3$}
\\\vspace*{-0.5em}
\textit{$^1$University of Salzburg}\\
\textit{$^2$University of Strathclyde}\\
\textit{$^3$Federal Reserve Bank of Cleveland}
}
\date{}


\def\equationautorefname~#1\null{%
  Eq.~(#1)\null
}
\def\equationautorefname~#1\null{
Eq.~(#1)\null
}

\begin{document}
\maketitle\thispagestyle{empty}\normalsize\vspace*{-2em}\small

\begin{center}
\begin{minipage}{0.8\textwidth}
\noindent\small \textbf{Abstract.} In light of widespread evidence of parameter instability in macroeconomic models, many time-varying parameter (TVP) models have been proposed.  This paper proposes a nonparametric TVP-VAR model using Bayesian additive regression trees (BART) that models the TVPs as an unknown function of effect modifiers. The novelty of this model arises from the fact that the law of motion driving the parameters is treated nonparametrically. This leads to great flexibility in the nature and extent of parameter change, both in the conditional mean and in the conditional variance. Parsimony is achieved through adopting nonparametric factor structures and use of shrinkage priors. In an application to US macroeconomic data, we illustrate the use of our model in tracking both the evolving nature of the Phillips curve and how the effects of business cycle shocks on inflation measures vary nonlinearly with changes in the effect modifiers. 
\\
\end{minipage}
\end{center}

\doublespacing\normalsize\renewcommand{\thepage}{\arabic{page}}
\newpage\normalsize
\section{Introduction}\label{sec:intro}

Econometric models used in macroeconomics have traditionally been linear and homoskedastic. Examples include vector autoregressions (VARs), dynamic factor models (DFMs), and linearized dynamic stochastic general equilibrium (DSGE) models, inter alia. However, in recent decades, there has been growing awareness of the empirical need to allow for parameter change and to relax homoskedasticity assumptions. 

Most of the models used to capture such parameter change allow for parameters to vary over time, but at a specific point in time the model remains linear. Structural break models, Markov switching models \citep[see][]{sims2006were},  and time-varying parameter (TVP) models are some prominent examples. TVP regressions and TVP-VARs, in particular, have been highly successful for structural macroeconomic analysis and forecasting \citep[see, for example,][]{dangl2012predictive, giannone2013, koop2013large, belmonte2014hierarchical, bitto2019achieving, korobilis2021high, huber2021inducing, hauzenberger2021fast}. In the TVP-VAR literature, parameters are assumed to evolve according to random walks or autoregressive processes.\footnote{It is also common to assume these random walks are independent of one another. But this assumption can translate into overfitting, since coefficients often feature substantial co-movement \citep[see][]{stevanovic2016common, chan2020reducing}.}
This also holds true for stochastic volatility processes, the addition of which is the most common way of relaxing the homoskedasticity assumption.  


A drawback of these parametric models of parameter change is that they risk mis-specification. That is, there is uncertainty about the specific law of motion driving changes in coefficients and volatilities. This not only includes how coefficients (volatilities) evolve over time but also whether changes in the law of motion depend on other observed factors (resulting in nonlinear interactions). Since policy makers often have a keen interest in how structural factors affect not only observed quantities (such as output or inflation) but also latent quantities that are often highly nonlinear functions of the reduced-form parameters of a TVP-VAR (such as the long-run unconditional mean, impulse response functions, or predictive distributions), this ignorance of possible relations between the parameters of a model and additional covariates is not merely a statistical problem but also has substantial practical relevance. For instance, a policy maker might want to know whether the effectiveness of policy measures depend on the state of the business cycle or whether macroeconomic trends (such as trend inflation or the equilibrium interest rate) depend dynamically on other factors.

The model developed in this paper allows for flexibility in how the parameters and the error volatilities in a VAR evolve over time. The model also allows for a (possibly) nonlinear relationship between a set of covariates, labeled effect modifiers, and the reduced-form parameters of the VAR. As will be explained, this simplifies use and interpretation of the model. Our model builds on those in two recent papers, \cite{deshpande2020vcbart} and \cite{coulombe2020macroeconomy}, which also model the dynamic evolution of parameters in a nonparametric manner.\footnote{\cite{fischer2022TVP} propose a parametric alternative that controls for uncertainty with respect to the form of the state evolution equation of a TVP-VAR.} Both of these papers --- that unlike this paper focus on univariate models --- assume that each coefficient has its own nonparametric law of motion. \cite{deshpande2020vcbart} approximate the law of motion using Bayesian additive regression trees \cite[BART,][]{chipman2010bart}, while \cite{coulombe2020macroeconomy} uses random forests. However, this flexibility might translate into overfitting and a lack of scalability to high dimensions.  Given our interest in (potentially) large-scale TVP-VARs with many more parameters, we address these issues by introducing further restrictions and Bayesian shrinkage priors so as both to lessen the risk of overfitting and to maintain computational tractability. 

In particular, the main contribution of the paper is the development of a flexible TVP-VAR that has several key model features that are important for inference. First, it allows for any time-variation in the parameters to be driven by a low number of latent factors. Second, it allows for factor stochastic volatility in the reduced-form VAR shocks. Third, it is nonparametric along several key dimensions: latent factors which define the law of motion of parameters (including both the TVPs and the error variance-covariance matrix) follow independent BART models. These BART models, in turn, depend on additional covariates that can be exogenous, endogenous, or latent. Novel shrinkage priors control for overfitting and allow for selection of an appropriate number of BART models. 

One advantage of our TVP-VAR, also noted by \cite{coulombe2020macroeconomy} in the univariate case, is that, by giving a nonparametric treatment to the parameters rather than the variables, our model remains conditionally linear in the parameters. This helps avoid the black-box nature of other nonparametric techniques, such as BART, since it facilitates interpretation of the model. Since we consider TVP-VARs and use them to carry out impulse response analysis, we can, for example, use our model to ask questions such as how impulse responses to structural shocks change if the effect modifiers are altered. This permits scenario analysis that is otherwise not possible (even with fully fledged nonparametric models) or strongly depends on the assumed relationship between the parameters and the effect modifiers.\footnote{Particular examples are \cite{hubrich2015financial}, \cite{ aastveit2017economic}, \cite{caggiano2017estimating}, \cite{alessandri2019financial}, and \cite{hauzenberger2021effectiveness}.} Our model remains agnostic on this relationship and thus can capture nonlinearities of arbitrary form. In addition, the heteroskedastic factor structure facilitates structural identification of the VAR model \citep[see][]{korobilis2022new, chan2022large}. 

We also devise a new efficient and scalable Markov chain Monte Carlo (MCMC) algorithm. The key features of our algorithm are that we simulate the trees of the TVPs marginally and thus avoid mixing issues that arise if one is sampled conditionally on the other. The factor model assumed for the shocks not only ensures parsimony but also allows us to exploit computational gains by rendering the different equations of the model conditionally independent. Hence, we can use fast equation-by-equation updating.

To illustrate use of the model, we revisit the debate on the possibly evolving nature of the Phillips curve in the US. Given heteroskedasticity, we first identify the error volatility factors up to sign and scale \citep[][]{chan2022large}, and then isolate a business cycle shock in a narrative fashion, by assuming that it is the factor that explains the largest share of variation in the reduced-form VAR residuals to output and unemployment variations during recessionary periods. Using this identified shock, we investigate the dynamic reactions of a panel of macroeconomic quantities with a particular focus on prices. Considering how different effect modifiers impact the posterior distributions of the impulse responses reveals  that the effects of business cycle shocks vary nonlinearly with uncertainty and according to whether the economy is in a recessionary regime. 

The remainder of the paper is structured as follows. Section \ref{sec:econ} introduces the general econometric framework. This section includes information on the likelihood function and our nonparametric treatment of the TVPs. Section \ref{sec:bayes} introduces our prior setup and discusses our posterior simulation algorithm, while Section \ref{sec: empwork} illustrates our techniques using US data. The final section concludes.
 
\section{Nonparametric time-varying parameter VARs}\label{sec:econ}
In this section we develop our flexible econometric model. Our goal is to model the evolution of an $M \times 1$-vector of macroeconomic time series, which we denote by $\{\bm y_t\}_{t=1}^T$. The elements in $\bm y_t$ might feature structural breaks, changing cross-variable dependencies, and/or different persistence behavior over time. In addition, it could be that the number of time series $M$ is large. Our framework will be capable of simultaneously handling many time series that might feature complex dynamics.
\subsection{A standard time-varying parameter model}
We assume that $\bm y_t$ evolves according to a VAR model with drifting parameters. Our TVP-VAR with $P$ lags is given by:
\begin{equation}
\bm y_t = \sum_{p = 1}^{P} \left(\bm A_{p} + \bm B_{pt}\right) \bm y_{t-p} + \bm \epsilon_t, \quad \bm \epsilon_t \sim \mathcal{N}(\bm 0_M, \bm \Omega_t) \label{eq:VAR}. 
\end{equation}

Note that we have written the TVP-VAR as involving $\bm A_{p}$ for $p=1,..,P$,  which are $M \times M$-matrices of constant coefficients, and $\bm B_{pt}$ for $p=1,..,P$,  which are $M \times M$-matrices of TVPs capturing deviations from the constant part of the model. $ \bm \epsilon_t$ is an $M \times 1$-vector of Gaussian shocks with mean zero and time-varying variance-covariance matrix $\bm \Omega_t$.  

There is substantial evidence that macroeconomic data are driven by a small set of fundamental shocks \citep{bai2007determining}. We incorporate this insight into our model by assuming that $\bm \epsilon_t$ features a common factor structure: 
\begin{equation}
\bm \epsilon_t = \bm \Gamma \bm q_t + \bm \varepsilon_t\quad  \Leftrightarrow \quad  \bm \Omega_t = \bm \Gamma \bm R_t \bm \Gamma' + \bm \Sigma_t, \label{eq: factordecomposition}
\end{equation}
where $\bm q_t$ is a $Q_q ( \ll M) \times 1$-vector of Gaussian-distributed factors with mean zero and diagonal time-varying variance-covariance matrix $\bm R_t = \text{diag}(r_{1t}, \dots, r_{Q_q t})$. $\bm \Gamma = (\bm \gamma_1, \dots, \bm\gamma_{Q_q})$ refers to a $M \times Q_q$-matrix of factor loadings, while $\bm \varepsilon_t = (\varepsilon_{1t}, \dots, \varepsilon_{Mt})'$ is an $M \times 1$-vector of Gaussian idiosyncratic shocks with mean zero and diagonal variance-covariance matrix $\bm \Sigma_t = \text{diag}(\sigma_{1t}^2, \dots, \sigma_{Mt}^2)$. One key observation is that, conditional on the factors $\bm q_t$, the shocks $\bm \varepsilon_t$ are independent and equation-by-equation estimation is possible. This factor structure has been used in other papers to facilitate estimation of large VARs \citep{kastner2020sparse, chan2021comparing, clark2021tail} and, in addition, this structure can also facilitate identification of the factors as structural VAR disturbances \citep{korobilis2022new,chan2022large}. Moreover, in contrast to VAR-based estimation using a Cholesky decomposition of the error covariances, another convenient feature of our model is that it is invariant to how the variables are ordered in $\bm y_t$  \citep[for a formal argument, see, for example,][]{chan2022large}.

It is worth stressing that the factor model, without additional restrictions, is not point identified.  This is because the factors and the loadings enter the likelihood in product form and are, thus, not invariant to rotation, column  and sign switching. Point identification can then be achieved by introducing restrictions on the loadings. A standard restriction assumes that the first $Q_q \times Q_q$ leading matrix of $\bm \Gamma$ is lower uni-triangular. This immediately implies that the resulting estimates will depend on the ordering of the elements in $\bm y_t$, a property that we would like to avoid.  Hence, in what follows we do not impose identification restrictions on our factor model during MCMC estimation. Results in \cite{chan2022large} suggest that the decomposition in \autoref{eq: factordecomposition} is identified up to column and sign switching.  Since our focus is on impulse responses to changes in particular elements in $\bm q_t$, we tackle column and sign switching ex-post. \cite{kaufmann2019bayesian} follow a similar strategy, post-processing the posterior factor draws to point-identify the factors/shocks.  We discuss this detail further in the empirical application below.


Up to this point we have remained silent on how the latent states (which include both the VAR coefficients and the time-varying elements of the error variances) evolve over time. In the next section, we introduce a flexible law of motion for the latent states.

\subsection{A nonparametric law of motion for the TVPs}\label{ssec:mean}
A standard assumption in the TVP-VAR literature is that the elements in $\bm B_{pt}$ $(p = 1, \dots, P)$ and $\bm R_t$ evolve according to simple parametric stochastic processes, most often random walks \citep{primiceri2005, cogley2005drifts, belmonte2014hierarchical, bitto2019achieving}. Assuming that the states evolve according to random walks introduces parsimony, because it implies a prior on the smoothness of the time variation in the coefficients. However, in turbulent periods, such as during the global financial crisis or the COVID-19 pandemic, it could be that parameters change rapidly and display sharp structural breaks. In such a case, a mixture model that models the evolution of the parameters as characterized by a low number of breaks \citep{sims2006were,koop2007estimation, kaufmann2015k} would be more appropriate. Another possibility is that parameter change could depend on exogenous \textit{effect modifiers}, translating into a specification with interaction effects. The nonparametric approach that we develop allows for all these possibilities.  

Although most TVP-VARs allow for each coefficient to have its own random walk process, this is probably too flexible. That is, it is an empirical regularity that there is a high degree of co-movement in the parameters. This motivates the inclusion of a factor structure in the TVPs (that is, allowing for the the process innovation variance-covariance matrix to be of reduced-rank). In the parametric TVP-VAR literature,  \cite{chan2020reducing} propose a model that assumes a factor structure on the TVPs and assumes that the factors driving the states evolve according to a random walk. \cite{fischer2022TVP} modify this approach by allowing for different forms of parameter change. This is achieved through including effect modifiers that can be either observed or latent.

In this paper, we do something similar, but we do it nonparametrically. That is, we assume there are a small number of latent nonlinear factors driving parameter change, which we estimate nonparametrically. In other words, we remain agnostic on the precise law of motion of the latent states, and let the data decide on the appropriate state dynamics, while achieving parsimony by introducing a factor structure to the TVPs.

We begin by writing the TVP-VAR in more compact form. Let $\bm x_t = (\bm y'_{t-1}, \dots, \bm y'_{t-p})'$ denote a $K = (MP) \times 1$-vector of covariates.  Moreover, let $\bm A = (\bm A_{1}, \dots, \bm A_{P})$ and $\bm B_t = (\bm B_{1t}, \dots, \bm B_{Pt})$ refer to $M \times K$-matrices that stack the VAR coefficients. Since our model, conditional on the latent factors, is a system of independent regression models we can focus on the $m^{th}$ equation of $\tilde{\bm y}_t = \bm y_t - \bm A \bm x_t$. This regression model can be expressed as: 
\begin{equation}
\tilde{y}_{mt} = \bm x_t'\bm \beta_{mt} + \bm q_t' \bm \gamma_m +\varepsilon_{mt}, \quad \bm q_t \sim \mathcal{N}(\bm 0_{Q_q}, R(\bm z_t)), \quad \varepsilon_{mt} \sim \mathcal{N}(0,\sigma^2_{mt}). \label{eq: eq_m}
\end{equation}
Here, $\bm \beta_{mt}$ and $\bm \gamma_m$ refer to the $m^{th}$ rows of $\bm B_t$ and $\bm \Gamma$, respectively.  The factors $\bm q_t$ arise from a Gaussian distribution with variance $\bm R_t = R(\bm z_t) = \text{diag}(r_{1}(\bm z_t), \dots, r_{Q_q}(\bm z_t))$ where $r_s: \mathbb{R}^N \to \mathbb{R}^+$ is an unknown function. Notice that the error variances depend on a set of effect modifiers in $\bm z_t$. 

We assume that $\bm \beta_{mt}$ evolves according to:
\begin{equation}
    \bm \beta_{mt} = \bm \Lambda_m  F_m(\bm z_t) + \bm \eta_{mt}, \label{eq:beta_state}
\end{equation}
where $F_m(\bm z_t) = (f_{m1}(\bm z_t), \dots, f_{mQ_\beta}(\bm z_t))'$ denotes an unknown function with $Q_\beta$ components $f_{mq}: \mathbb{R}^N \to \mathbb{R}$,  $\bm \Lambda_m$ is a $K \times Q_\beta$-matrix of factor loadings, and $Q_\beta$ is the number of latent factors that drive the TVPs. In addition, we assume that  $\bm \eta_{mt} \sim \mathcal{N}\left(\bm 0_K, \bm V_m\right)$ is a vector of Gaussian shocks with $\bm V_m = \text{diag}(v_{m1}^2, \dots, v_{mK}^2)$ denoting the process innovation variances. 

Conditional on choosing an appropriate number of factors $Q_\beta$, this specification is extremely flexible. It allows for (potentially) nonlinear interactions between $\bm z_t$ and $\bm \beta_{mt}$ (and thus implicitly $\bm x_t$). If elements in $\bm \beta_{mt}$ do not depend on $\bm z_t$ the corresponding loadings are zero and time-variation can still be captured through the presence of the idiosyncratic shocks in $\bm \eta_{mt}$.\footnote{Our model can also be related to random coefficient models, see \cite{fruhwirth2004bayesian}.}  Notice that the variances in $\bm V_m$ also control the weight put on the nonlinear factor component. For instance, if the $j^{th}$ coefficient closely co-moves with the other coefficients in a nonlinear manner, $v_{mj}^2$ will be close to zero. 

To make this model operational we have to learn the functions $F$ and $R$ and decide on appropriate effect modifiers in $\bm z_t$. Our approach remains agnostic on the specific shape of both $F$ and $R$ and uses BART to estimate them.  The next sub-sections show how this is achieved. 

The choice of effect modifiers should depend on the application. The  modifiers could include exogenous regressors, deterministic functions of time,  lagged elements of $\bm y_t$, or latent quantities. If elements in $\bm z_t$ are endogenous and interest centers on higher-order impulse responses or multi-step-ahead predictive densities, one could either set up a separate law of motion for $\bm z_t$ or introduce hard restrictions on how the $\bm z_t's$ are expected to evolve over the forecast/impulse response horizon. In our empirical application below, we follow the latter approach, not only for simplicity, but because we are interested in how the dynamic reactions of $\bm y_t$ to shocks depend on the elements in $\bm z_t$ taking on certain values.\footnote{This would resemble common practice in, e.g., threshold or Markov switching models that condition on the prevailing regime when computing impulse responses.} This enables us to answer what-if questions, such as, ``How would inflation react to business cycle shocks if uncertainty is (and remains) high?", or, ``How do price reactions to business cycle movements change if the population becomes increasingly over-aged?" 

Another interesting possibility would be to set $\bm z_t= \bm x_t$. In this case, however, interpretation becomes more difficult since the model then becomes nonlinear in $\bm y_t$. This would then necessitate the use of generalized impulse responses \citep{koop1996impulse} to carry out dynamic analysis. With our application seeking to characterize features of the US business cycle, we choose effect modifiers that either slowly evolve independently of the business cycle, like factors related to the age-composition of the population,  or binary variables (such as recession indicators), or other variables not included in $\bm y_t$. One such variable we consider is the uncertainty measure proposed in \cite{jurado2015measuring} which can be interpreted as a proxy of (unobserved) macroeconomic uncertainty. 

Another strategy to selecting the elements of  $\bm z_t$ would be to entertain a large set of potential effect modifiers, and then use regularization techniques. As we will describe below, our approach is capable of handling all these cases without additional modification.

\subsection{Learning the unknown functions using BART}

We approximate  each function $f_{mq}$ through a sum-of-trees model \citep{chipman2010bart}:
\begin{equation}
f_{mq} \approx g(\bm Z| \mathcal{T}^{\beta}_{mq}, \bm \mu_{mq}) = \sum_{s=1}^{S_\beta} u \left(\bm Z| \mathcal{T}^{\beta}_{mq,s}, \bm \mu_{mq,s}\right), \label{eq:BART}
\end{equation}
with the $T \times N$-matrix $\bm Z$ having a typical $t^{th}$ row $\bm z'_t$ and $u$ being a regression tree function that depends on a tree structure, $\mathcal{T}^{\beta}_{mq} = \{\mathcal{T}^\beta_{mq,1}, \dots, \mathcal{T}^\beta_{mq, S_\beta}\}$, which is a sequence of disjoint sets that partition the input space and a vector of terminal node parameters $\bm \mu_{mq} = \{\bm \mu_{mq,1}, \dots, \bm \mu_{mq,S_\beta}\}$ of dimension $b_{mq}$.  These partitions are driven by splitting rules of the form $z_{jt} \le c_j$ or $z_{jt} > c_j$, with $z_{jt}$ denoting the $j^{th}$ element of $\bm z_t$ and $c_j$ being a threshold parameter. Moreover, $S_\beta$ is the number of trees used to approximate each of the functions (factors) $f_{mq}$. \autoref{eq:BART} is a standard BART model. To avoid issues associated with overfitting when $S_\beta$ is large, \cite{chipman2010bart} propose using a regularization prior to force the trees to take a particularly simple form and thus explain only a small fraction of the variation of the response variable. Adding together many simple trees (weak learners) has been found to work better than working with a single more complicated tree. We follow such an approach in this paper. 

Plugging (\ref{eq:BART}) into (\ref{eq:beta_state}) yields our state equation:
\begin{equation}
\bm \beta_{mt} = \sum_{q=1}^{Q_\beta} \bm \lambda_{mq} {g(\bm z_t| \mathcal{T}^{\beta}_{mq}, \bm \mu_{mq})} + \bm \eta_{mt}, \label{eq:beta_state2}
\end{equation}
with $\bm \lambda_{mq}$ denoting the $q^{th}$ column of $\bm \Lambda_m$. This shows that we combine $Q_\beta$ BART models (each used to approximate one of the $Q_\beta$ functions). Setting $Q_\beta=1$ implies that all coefficients are driven by a single factor (if $\bm \lambda_{mq} \neq \bm 0_K$), while when setting $Q_\beta=K$ we obtain a model closely related to the one proposed in \cite{deshpande2020vcbart} and \cite{coulombe2020macroeconomy}. Since the latter specification, in light of large $K$, does not scale well to high dimensions we will focus on the case where $K \gg Q_\beta$, which frequently arises in the analysis of large TVP-VAR models. We will call models that assume this nonparametric factor form for the conditional mean, TVP-FBART. Ahead of our main empirical application and to help the reader further understand our model, Sub-section \ref{app:illust} in the Online Appendix provides a toy empirical example to illustrate how BART can be used to approximate TVPs.

\subsection{Flexible heteroskedasticity specifications}\label{ssec:var}
Recall that our model also assumes that the shocks $\bm \epsilon_t$ feature a factor structure. We will again approximate the factor-specific functions, $r_s(\bm z_t)$, in $\bm R_t$   with BART. More precisely, our approach can be interpreted as a variant of heteroskedastic BART \citep[heteroBART, see ][]{pratola2020heteroscedastic}. heteroBART is a multiplicative version of BART and assumes that the trees enter the model in product form. In this paper, we follow \cite{clark2021tail} and linearize the model so that standard BART techniques can be used.

Let the $s^{th}$ element of $\bm q_t$ be given by:
\begin{equation}
q_{st} = \exp\left(\sum_{d=1}^{S_q} u(\bm z_t| \mathcal{T}^{q}_{sd}, \bm \pi_{sd})/2\right) \times  \xi_{st}, \quad \xi_{st} \sim \mathcal{N}(0, 1), \label{eq: vola_q}
\end{equation}
with $S_q$ being the number of trees used to approximate the variance functions, where $\mathcal{T}^{q}_{sd}$ and $\bm \pi_{sd}$ denote the corresponding tree structures and terminal node parameters, respectively. To render \autoref{eq: vola_q} linear we square it and take logs. This yields a linear equation with shocks that are log-$\chi^2$ distributed with one degree of freedom, a distribution which can be well approximated using  a ten-component mixture approximation \citep[see][]{omori2007stochastic}:
\begin{equation}
\tilde{q}_{st} = \log (q_{st}^2)= \sum_{d=1}^{S_q} u(\bm z_t| \mathcal{T}^{q}_{sd}, \bm \pi_{sd}) + \tilde{\xi}_{st},\quad \tilde{\xi}_{st} = \log (\xi_{st}^2)\sim \sum_{i=1}^{10} w_i \mathcal{N}(m_i, \mathfrak{r}_i^2).
\end{equation}
Here, $m_i$, $w_i$, and $\mathfrak{r}_i^2$ are fixed numbers defining the mixture components taken from Table 1 in \cite{omori2007stochastic}. 

This specification of heteroBART implies that the factor volatilities are allowed to change rapidly, but can also move more gradually. This feature might pay off during recessions, where large jumps in error volatilities are common. Traditional stochastic volatility models will be unable to match this pattern, since they assume that the log-volatilities evolve according to a stochastic process that translates into a more gradual evolution of the error variances. Such behavior is warranted if the trend movement in volatility is persistent (such as during the Great Moderation). For heteroBART, matching slowly evolving trends is also possible but considerably harder. To allow for smoothly evolving stochastic trends we combine heteroBART with a standard stochastic volatility model in the measurement errors (that is, the elements in $\bm \Sigma_t$). The combination between a parametric law of motion for $\log(\sigma^2_{mt})$  and $q_{st}$ allows for rich dynamics in terms of $\bm \Omega_t$.\footnote{Another option to capture smoothly varying trends with heteroBART would be through the specification of a latent component which enters $\bm z_t$. But this would require nonlinear filtering algorithms or linear approximations (which can fail in certain environments) such as the ones proposed in \cite{huber2020nowcasting}.}
We will use the abbreviation FHB (using a factor structure involving heteroBART) for models which adopt this specification. Thus, our most general model is TVP-FBART-FHB. 


\subsection{Summary of key model features}\label{ssec:sum}
The model described in the previous sub-sections is very flexible and nests a wide variety of competing models.  In this sub-section, we first   summarize key model features and then discuss how our model is related to alternative models commonly used in the literature. 

Flexible machine learning techniques such as BART have the shortcoming that interpretability is difficult. As noted, for example in \cite{coulombe2020macroeconomy}, using regression trees to model the parameters of a TVP regression allows for flexibility but also maintains simplicity of interpretation. In our case, once we have learned the TVPs and the functions driving them using BART, interpretation of the model works analogously to a standard TVP-VAR model. Hence, one can compute functions of the parameters such as impulse responses, forecast error variance or historical decompositions, and conditional forecasts using standard techniques. This constitutes a big advantage of our approach relative to models such as the one proposed in \cite{huber2022inference}. Computation of (generalized) impulse response function in traditional BART-based VAR models is much more involved, as the model remains nonlinear.\footnote{\cite{koop1996impulse} discuss how to compute generalized impulse response functions in nonlinear multivariate models.} 

The previous paragraph is related to the effect that shocks might have on $\bm y_t$. Since the effect modifiers influence $\bm y_t$ indirectly through the BART modeling of the TVPs, we can also assess how $\bm z_t$ affects $\bm y_t$. This can be easily achieved in our framework since one can compute different realizations of the TVPs for different configurations of $\bm z_t$. Doing so allows us to study how (higher-order) interaction effects, which might take an unknown form, impact quantities such as forecast distributions, impulse responses, or even long-run trends such as the (time-varying) unconditional mean of the TVP-VAR. We will illustrate these features in our empirical work that follows in Section \ref{sec: empwork} below.

Apart from the ease of interpretation and the additional inferential possibilities, our model, for appropriately chosen values of $Q_\beta, Q_q, S_\beta$, and $S_q$, provides a great deal of flexibility when it comes to capturing different forms of parameter change. While our aim is to introduce as few restrictions on the state evolution as possible, we can nevertheless control the dynamics of the TVPs by choosing appropriate values of $S_\beta$ and $S_q$. In principle, larger values of $S_\beta$ and $S_q$ are consistent with smooth law of motions of the parameters, whereas smaller values imply parameter dynamics closer to the ones generated by a structural break model. An extreme case of our model would set $Q_\beta = S_\beta = 1$. This specification would imply that parameters follow a single regression tree and are proportional to each other. In our empirical work we will explore the sensitivity of results by varying these parameters.

\section{Bayesian inference}\label{sec:bayes}
\subsection{The prior}
We start our discussion with the priors relating to the regression trees and the process innovation variances. The remaining priors are relatively standard and a discussion can be found in Section \ref{app:A} in the Online Appendix.

\cite{chipman1998bayesian} and \cite{chipman2010bart} specify a tree-generating stochastic process on the tree structures, $\mathcal{T}^{\beta}_{mq, s}$ and $\mathcal{T}^{q}_{sd}$. Our approach is similar, but specifies the prior such that the probability of growing more complex trees decreases with the number of factors $\nu =1,\dots, Q_j$ for  $j \in \{\beta, q\}$.  This process is designed to penalize complex trees and consists of three features:
\begin{enumerate}[leftmargin=*]
    \item A decreasing probability that a node is non-terminal. Let $\mathfrak{n}=0, 1, \dots$ denote a particular node at depth $\mathfrak{n}$. We model the probability that this node is non-terminal as follows:
    \begin{equation*}
        \alpha^\nu (1+\mathfrak{n})^{-\zeta^\nu},
    \end{equation*}
    where $\alpha$ is between $0$ and $1$ and $\zeta > 1$. This implies that $\alpha$ acts as a base parameter and $\zeta$ penalizes more complex trees by shrinking the probability that a given node is non-terminal for higher-order nodes. A typical choice that works well for many datasets  is $\alpha = 0.95$ and $\zeta =2$ \citep[see][]{chipman2010bart}. Notice that the base probability decreases in $\nu$, implying that for large values of $\nu$ the probability of forking a new branch of a tree decreases substantially. This effect is complemented by the shrinkage parameter $\zeta^\nu$ which grows rapidly in $\nu$. By setting $\nu = q$, this prior effectively allows the modeler to select $Q_\beta$ by forcing the functions $f_{mq}$, for large $q$, towards a constant function (implying no effect on coefficient dynamics).  This prior is used for $\mathcal{T}^{\beta}_{mq,s}$. For $\mathcal{T}^{q}_{sd}$, we set $\nu=1$ (regardless of $s$) and thus use the benchmark prior of \cite{chipman2010bart}.
    
    \item A prior distribution on the splitting variables in $\bm z_t$. In the absence of strong prior information we follow much of the literature and use a discrete uniform prior on the elements in $\bm z_t$. Hence, at every node, every variable in $\bm z_t$ is equally likely to be used to split up the input space.
    \item A prior on the thresholds $c_j$ within a given splitting rule is assumed to be uniformly distributed. Similar to the prior on the splitting variables, this specification remains agnostic on the precise values that the thresholds may take.
\end{enumerate}
This prior encourages smaller trees and is thus consistent with the notion that each individual tree is a ``weak learner," but the composite model is capable of capturing complex dynamics in the parameters.

The prior on the terminal node parameters is Gaussian. Following \cite{chipman2010bart}, we scale the data such that the dependent variable is between $-0.5$ and $0.5$ and our prior covers this range. Let $\mu_{mq, ij}$ denote the $j^{th}$ element of $\bm \mu_{mq, i}$ and $\pi_{sd,j}$ the $j^{th}$ element of $\bm \pi_{sd}$. The prior for the respective element is then given by:
\begin{equation*}
    \mu_{mq, ij} \sim \mathcal{N}\left(0, \frac{1}{2 \kappa S_\beta}\right) \quad \text{and} \quad  \pi_{sd, j} \sim \mathcal{N}\left(0, \frac{1}{2 \kappa S_q}\right).
\end{equation*}

Here, $\kappa$ is a parameter that controls the prior variance. Shrinkage is introduced by increasingly forcing $\mu_{mq, ij}$ ($\pi_{sd, j}$) towards zero if $S_\beta$ ($S_q$) is large. Since $S_j$ ($j \in \{\beta, q\}$) is typically between 50 and 200, this prior is the second ingredient of BART used to capture the notion that each tree explains only a small amount of variation in $\bm \beta_t$ (and $q_{st}$).

On the different elements of $\bm V_m$, several priors are possible. The simple conjugate inverse Gamma prior can be used. This prior, however, has implications for our model, since it rules out values of $v^2_{mj}$ very close to zero. Hence, it would artificially push the likelihood away from the factor part in \autoref{eq:beta_state}. We follow recommendations in \cite{fs_wagner} and use a prior that introduces shrinkage on $\bm V_m$. Our prior assumes that $v^2_{mj}$ arises from a Gamma distribution:
\begin{equation*}
    v^2_{mj} \sim \mathcal{G}\left(\frac{1}{2}, \frac{1}{2 B_v}\right) \quad \Leftrightarrow \quad \pm v_{mj} \sim \mathcal{N}(0, B_v),
\end{equation*}
with $B_v$ being a scalar hyperparameter that controls the amount of shrinkage towards a factor structure in the TVPs. Since there exists strong evidence that the TVPs feature a factor structure, we set $B_v=0.01$ to have a tight prior on the idiosyncratic deviations of the TVPs from the common factor structure. 

\subsection{Markov chain Monte Carlo sampling} \label{sec:MCMC}
We sample from the joint posterior distribution of the model by using an MCMC algorithm that, conditional on the latent factors, simulates the coefficients and latent states for each equation separately. Since for some of the steps in the sampler we integrate out other parameters the precise ordering of the steps of the MCMC algorithm is important to simulate from the correct stationary distribution.
Our algorithm cycles between the following steps. 
For each equation $m=1,\dots, M$:
\begin{enumerate}[leftmargin=*]
    \item \textbf{Sampling the trees.} We sample the regression trees associated with the VAR coefficients marginally of the TVPs and conditional on the remaining parameters and latent states. Similar to \cite{chipman2010bart}, the trees are simulated on a tree-by-tree basis. Specifically, to sample the $q^{th}$ tree conditional on the other trees we first integrate out the TVPs of the $m^{th}$ equation by plugging the state equation in \autoref{eq:beta_state2} into the observation equation in \autoref{eq: eq_m}. Subtracting $\bm q_t'\bm \gamma_m$ from $\tilde{y}_{mt}$ and carrying out some algebraic manipulations yields:
    \begin{equation*}
       \frac{\tilde{y}^*_{mt} - \sum_{j \neq q} \tilde{x}_{mj,t} g(\bm z_t|\mathcal{T}^\beta_{mj}, \bm \mu_{mj})}{\tilde{x}_{mq,t} }   = g(\bm z_t|\mathcal{T}^\beta_{mq}, \bm \mu_{mq}) + \tilde{\varsigma}_{mt},
    \end{equation*}
    with $\tilde{y}^*_{mt}= \tilde{y}_{mt} - \bm q_t'\bm \gamma_m,\tilde{\bm x}_{mt} = \bm x_t' \bm \Lambda_m$ and $\tilde{x}_{mq,t}$ denoting the $q^{th}$ element of $\tilde{\bm x}_{mt}$. Moreover, $\tilde{\varsigma}_{mt} \sim \mathcal{N}\left(0, \frac{\bm x_t' \bm V_m \bm x_t + \sigma^2_{mt}}{\tilde{x}^2_{mq,t}}\right)$ refers to a period-specific (independent) white noise shock. This is a nonparametric regression model and the  Metropolis Hastings algorithm proposed in \cite{chipman1998bayesian} can be used to simulate the tree structures.
    
    \item \textbf{Sampling the terminal node parameters.} Conditional on the tree structures we can obtain the terminal node parameters by sampling from univariate Gaussian posterior distributions. The corresponding moments take particularly simple forms since, conditional on a tree structure that allocates observations to a specific terminal node, the posterior resembles the one of a simple intercept model under a conjugate Gaussian prior.
    
    \item \textbf{Sampling the coefficient factor loadings.} We sample the factor loadings in $\bm \Lambda_m$ conditional on the estimated trees, the process innovation variances in $\bm V_m$, the latent factors $\bm q_t$ but marginally of the TVPs. This can be achieved as follows. The observation equation, after integrating out the TVPs, can be written as:
    \begin{equation*}
        \tilde{y}^*_{mt} = \bm x'_t \bm \Lambda_m F_m(\bm z_t) + \varsigma_{mt}, \quad \varsigma_{mt} \sim \mathcal{N}\left(0, \bm x_t' \bm V_m \bm x_t + \sigma^2_{mt}\right).
    \end{equation*}
    Notice that $\bm x'_t \bm \Lambda_m F_m(\bm z_t)$ equals $(F'_m(\bm z_t) \otimes \bm x'_t) \text{vec}(\bm \Lambda_m)$, implying a standard multivariate regression model. Hence, under the Horseshoe prior, $\text{vec}(\bm \Lambda_m)$ follows a multivariate Gaussian posterior distribution with posterior mean and variance taking standard forms.  
    
    
    \item \textbf{Sampling the TVPs.} To sample the full history of $\{\bm \beta_{mt}\}_{t=1}^T$, we exploit the static representation of the model in \autoref{eq: eq_m}. Let $\tilde{\bm y}^{\bullet}_{m}$ denote a $T \times 1$-vector with typical element $\tilde{y}^{\bullet}_{mt} = (\tilde{y}_{mt} - \bm q_t'\bm \gamma_m)/\sigma_{mt}$ and ${\bm  W}_m= \text{bdiag}(\bm x'_1/\sigma_{m1}, \dots, \bm x'_T/\sigma_{mT})$ is a $T \times TK$-dimensional block diagonal matrix with the normalized $\bm x'_t$'s along its main diagonal. The corresponding static representation of the model is:
    \begin{equation*}
        \tilde{\bm y}^{\bullet}_{m} = {\bm  W}_m \bm \beta_m + \tilde{\bm \varepsilon}_{m}, \quad \tilde{\bm \varepsilon}_{m} \sim \mathcal{N}(\bm 0_T, \bm I_T),
    \end{equation*}
    with $\bm \beta_m = (\bm \beta'_{m1}, \dots, \bm \beta'_{mT})'$ and $\tilde{\bm \varepsilon}_{m} = (\varepsilon_{m1}, \dots, \varepsilon_{mT})'$. The corresponding posterior distribution of $\bm \beta_m$ is a $TK$-dimensional Gaussian distribution with a block-diagonal posterior covariance matrix:
    \begin{equation*}
    \begin{aligned}
        &\bm \beta_m | \bullet \sim \mathcal{N}(\overline{\bm \beta}_m, \overline{\bm V}_{\bm \beta_m}), \quad \text{where}\\
        &\overline{\bm V}_{\bm \beta_m} = \left({\bm  W}_m' {\bm  W}_m + (\bm I_T \otimes \bm V^{-1}_m)\right)^{-1} \quad \text{and} \quad 
        \overline{\bm \beta}_m = \overline{\bm V}_{\bm \beta_m} \left({\bm  W}_m'\tilde{\bm y}^{\bullet}_{m} + (\bm I_T \otimes \bm V^{-1}_m) \underline{\bm \beta}_m\right).
    \end{aligned}
    \end{equation*}
    The $TK$-dimensional prior mean vector is given by $\underline{\bm \beta}_m = \left((\bm \Lambda_m F_m(\bm z_1))', \dots, (\bm \Lambda_m F_m(\bm z_T))'\right)'$. \cite{hauzenberger2021fast} provide fast algorithms to sample from this posterior distribution which exploit the fact that the rank of ${\bm  W}_m' {\bm  W}_m$ is $T$.  
    
    \item \textbf{Sampling the process innovation variances.} The posterior distribution of the process innovation variances $v_{mj}^2$ follows a generalized inverse Gaussian (GIG) distribution:
    \begin{equation*}
        v_{mj}^2|\bullet \sim \mathcal{GIG}\left(1/2 - T/2, \sum_{t=1}^T \eta^2_{mj,t}, (2 B_v)^{-1}\right).
    \end{equation*}
    Here, $\eta_{mj,t}$ denotes the $j^{th}$ element of $\bm \eta_{mt}.$
    
    \item \textbf{Sampling the time-invariant regression coefficients.} Conditional on the TVPs,  time-varying error variances, latent factors and loadings, the time-invariant regression coefficients can be obtained from standard multivariate Gaussian posterior distributions. More precisely, the $m^{th}$ row of $\bm A$, $\bm a_m$, arises from a $K$-dimensional Gaussian posterior:
    \begin{equation*}
    \begin{aligned}
        &\bm a_m|\bullet \sim \mathcal{N}(\overline{\bm a}_m, \overline{\bm V}_{a_m}), \quad \text{with moments}\\
        & \overline{\bm V}_{a_m} = (\bm X'\bm \Sigma_m^{-1} \bm X + \underline{\bm V}^{-1}_{a_m})^{-1} \quad \text{and} \quad 
        \overline{\bm a}_m = \overline{\bm V}_{a_m} \bm X' \bm \Sigma_m^{-1} \overline{\bm y}_m.
    \end{aligned}
    \end{equation*}
    Here, $\underline{\bm V}_{a_m}$ is the diagonal prior covariance matrix with typical element given by the Horseshoe prior described in the previous sub-section and $\overline{\bm y}_m$ is a $T \times 1$-vector with $t^{th}$ element $(y_{mt} - \bm x'_t \bm \beta_{mt} - \bm q_t'\bm \gamma_m)$.
    
     \item \textbf{Sampling the factor loadings $\bm \gamma_m$.} The factor loadings $\bm \gamma_m$ can be straightforwardly obtained by estimating a regression model with covariates $\bm q_t$, response variable $\tilde{y}_{mt} - \bm x_t' \bm \beta_{mt}$, and heteroskedastic shocks with variances $\bm \Sigma_m$. The corresponding posterior distribution of $\bm \gamma_m$ is Gaussian and the moments take a form similar to the ones in Step 6.
    
    \item \textbf{Sampling the hyperparameters associated with the  prior on $\bm a_m = (a_{1m}, \dots, a_{Km})'$.} To sample the diagonal elements of $[\underline{\bm V}_{a_m}]_{ii} = \varrho^2_{i, a_m} \varpi^2_{a_m}$, with $\varrho_{i, a_m}$ and $\varpi_{a_m}$ denoting the $i^{th}$ local shrinkage parameter and the equation-specific global shrinkage parameter,  we use the efficient and simple-to-implement sampler proposed in \cite{makalic2015simple}. This sampler introduces two types of auxiliary random variables, $r_{i, a_m} \sim \mathcal{G}^{-1}(1/2, 1)$ and $n_{a_m} \sim \mathcal{G}^{-1}(1/2, 1)$, that have inverse Gamma priors. Simulating from the posterior of $\varrho^2_{i, a_m}$ and  $\varpi^2_{a_m}$ is then achieved by first simulating $r_{i, a_m}$ and $n_{a_m}$ from inverse Gamma distributions:
    \begin{equation*}
        r_{i, a_m}| \bullet \sim \mathcal{G}^{-1}\left(1, 1 + \frac{1}{\varrho^2_{i, a_m}}\right) \quad \text{and} \quad 
        n_{a_m}|\bullet \sim \mathcal{G}^{-1} \left(1, 1 +  \frac{1}{\varpi^2_{a_m}}\right).
    \end{equation*}
    Conditionally on a draw of $r_{i, a_m}, n_{a_m}$, the full conditional posterior of $\varrho_{i, a_m}^2$ and $\varpi_{a_m}^2$ is inverse Gamma as well:
    \begin{equation*}
        \varrho_{i, a_m}^2|\bullet \sim \mathcal{G}^{-1}\left(1, \frac{1}{r_{i, a_m}} + \frac{a^2_{im}}{2 \varpi^2_{a_m}}\right) \quad \text{and} \quad 
        \varpi^2_{a_m}|\bullet \sim  \mathcal{G}^{-1}\left(1, \frac{1}{n_{a_m}} + \frac{1}{2} \sum_{i=1}^K \frac{a^2_{im}}{\varrho_{i, a_m}^2} \right).
    \end{equation*}
    
    \item \textbf{Sampling the hyperparameters associated with the  prior on $\bm \Lambda_m$.} This step closely mirrors Step 7, with the relevant quantities being replaced by the corresponding elements in $\bm \Lambda_m$.
    \item \textbf{Sampling the  latent log-volatilities and the parameters of the state equation.} We sample the log-volatilities, $\log (\sigma^2_{m1}), \dots, \log (\sigma^2_{mT})$, and the parameters of the corresponding state equation (which involve the unconditional mean, the persistence parameter, and the variance of the shocks to the log-volatilities) using the efficient sampler devised in \cite{kastner2014ancillarity} and implemented in the \texttt{R} package stochvol \citep{kastner2016dealing}.
\end{enumerate}
The following quantities are not estimated in an equation-by-equation manner:
\begin{enumerate}[leftmargin=*]
 \setcounter{enumi}{10}
    \item \textbf{Sampling the latent factors $\{\bm q_t\}_{t=1}^T$.} The latent factors in $\bm q_t$ are simulated on a $t$-by-$t$ basis. This can be achieved by estimating $T$ separate regressions by regressing $\hat{\bm y}_t = \bm \Sigma_t^{-1/2} \left(\bm y_t - \sum_{p=1}^P (\bm A_p + \bm B_{pt}) \bm y_{t-p}\right)$ on $\hat{\bm X}_t =  \bm \Sigma_t^{-1/2} \bm \Gamma$ for all $t$. The corresponding time $t$ posterior of $\bm q_t$ is Gaussian:
    \begin{equation*}
        \bm q_t|\bullet \sim \mathcal{N}(\overline{\bm q}_t, \overline{\bm V}_{q_t}),
    \end{equation*}
    with covariance matrix and mean vector given by, respectively:
    \begin{equation*}
        \overline{\bm V}_{q_t} = \left(\hat{\bm X}'_t \hat{\bm X}_t  + \bm R_t^{-1}\right)^{-1} \quad \text{and} \quad 
        \overline{\bm q}_t = \overline{\bm V}_{q_t} \hat{\bm X}'_t \hat{\bm y}_t.
    \end{equation*}
    \item \textbf{Sampling the hyperparameters of the Horseshoe associated with $\bm \Gamma$.} The Horseshoe shrinkage parameters on the factor loadings in $\bm \Gamma$ are simulated analogously to Step 8 of the algorithm. The main difference, however, is related to the fact that we specify $Q_q$ global shrinkage parameters (one for each column) and the corresponding full conditional posterior for $\varpi^2_{\Gamma_j}$ needs to be adjusted by summing only over the relevant local scales and parameters  associated with and in the $j^{th}$ column of $\bm \Gamma$, $\bm \Gamma_j$.
\end{enumerate}
Notice that steps (1) to (3) yield a draw from $p(\{\mathcal{T}^\beta_{mq}, \bm \mu_{mq}, \bm \lambda_{mq}\}_{q=1}^{Q_\beta}|\bm \bullet_{/\bm \beta_{mt}})$ where the notation $\bullet_{/\bm \beta_{mt}}$ indicates the remaining model parameters except the TVPs and the data. The TVPs are then simulated from $p(\{\bm \beta_{mt}\}_{t=1}^T|\bullet)$ where $\bullet$ means all other model parameters, latent quantities and the data. This step differs from the one used in \cite{deshpande2020vcbart} since we improve mixing by integrating out the TVPs. In principle, the loadings and trees can also be sampled conditionally on the TVPs but in cases where the loadings are very small substantial mixing issues arise.

We repeat this algorithm $15,000$ times and discard the first $5,000$ draws as burn-in.\footnote{To obtain $15,000$ draws, the actual computation time is about $124$ minutes, based on a MacBook Pro with an M1 8-core processor.} From a computational perspective, this algorithm is quite efficient. This is because the sampling step associated with the TVPs can be sped up enormously by exploiting the fact that $\bm W_m' \bm W_m$ is a block-diagonal matrix of rank $T$.

\section{Empirical application: Modeling inflation}\label{sec: empwork}
\subsection{Data overview and specification choices} We use the quarterly version of the \cite{mccracken2016fred} data set and focus on a sample ranging from  $1975$:Q$1$ to $2019$:Q$4$. In our empirical work, we aim to investigate how business cycle shocks impact a range of different price measures and whether these dynamic reactions depend on the effect modifiers. To this end, we follow \cite{del2020s} and estimate medium-sized VAR models that are rich in wage, price, and labor market measures. We consider $M=12$ endogenous variables, where $\bm y_t$ includes output growth, employment, unemployment, average weekly hours worked, personal consumption expenditure (PCE) inflation, PCE inflation excluding food and energy, (core) consumer price inflation, the GDP deflator, wage inflation, the federal funds rate, and  ten-year government bond yields to capture movements in treasury markets. But unlike \cite{del2020s}, we allow for nonlinear relationships between these variables and for these effects to vary over time. \cite{del2020s} accommodate temporal change only, by simply estimating their linear VAR model over two non-overlapping samples. 

As effect modifiers in $\bm z_t$, we consider five indicators that may affect the TVPs, and in turn the impulse response functions, in a nonlinear manner. Specifically, we consider the old-age dependency ratio, a financial globalization indicator, the (lagged) ex-post real rate, a binary recession indicator (taken from the NBER), and the economic uncertainty index proposed in \cite{jurado2015measuring}. Secular stagnation factors, such as a boost in financial globalization, the rising old-age dependency ratio, and a declining real rate, may affect the dynamics of business cycle phases in a nonlinear manner \citep{jones2022aging}. These factors have also been identified as one cause of the flattening of the Phillips curve  \citep{forbes2019inflation, forbes2021low}. The last two effect modifiers allow for possible structural breaks in recessionary and high uncertainty periods \citep[see, for example,][]{aastveit2017economic, alessandri2019financial}. 

Some of the effect modifiers are clearly endogenous and should depend on the other quantities of our model. This does not cause any issues for the validity of our econometric approach. However, when we focus on impulse responses it has the implication that $\bm z_t$ is not allowed to react to changes in $\bm y_t$. As discussed in Section \ref{sec:econ}, this is an assumption made for the sake of interpretability. The main implication is that impulse responses can be understood as being conditional on $\bm z_t$ remaining at the current level over the impulse response horizon. Since we are going to construct ``scenarios,'' based on assumptions about how $\bm z_t$ behaves, this restriction can be interpreted as similar in nature to conditional forecasts when the restricted variables are not located in $\bm y_t$ but in $\bm z_t$. If the researcher wishes to relax these assumptions, they can set up auxiliary models for $\bm z_t$, such that $\bm z_t$ is again a function of $\bm y_t$.

Table \ref{tab:data} in the Online Appendix provides additional information on the time series and associated data transformations used. All models we consider in this paper feature $p=5$ lags. In Sub-section \ref{app:model_fit} we assess how different model features impact model fit and compare our proposed model to standard models in the literature. This analysis evidences that our model generally captures the data well, often improving upon competitors commonly used in the literature. Based on the results in Table \ref{tab:WAIC}, we use the model that sets $Q_\beta=25$, $S_\beta=1$, $Q_q=3$, and $S_q=250$.

\subsection{Some features of our estimated model}\label{ssec:WAIC}
\begin{figure}[!htbp]
    \centering
    \caption{Explaining the dynamics in coefficients of the conditional mean.}
    \label{fig:insmpMEAN}
\begin{minipage}{0.49\textwidth}
\centering
(a) Total share of time-variation explained 
\end{minipage}
\begin{minipage}{0.49\textwidth}
\centering
(b) Number of tree splits
\end{minipage}
\begin{minipage}{0.49\textwidth}
\centering
\vspace*{5pt}
\includegraphics[width=1\textwidth,keepaspectratio]{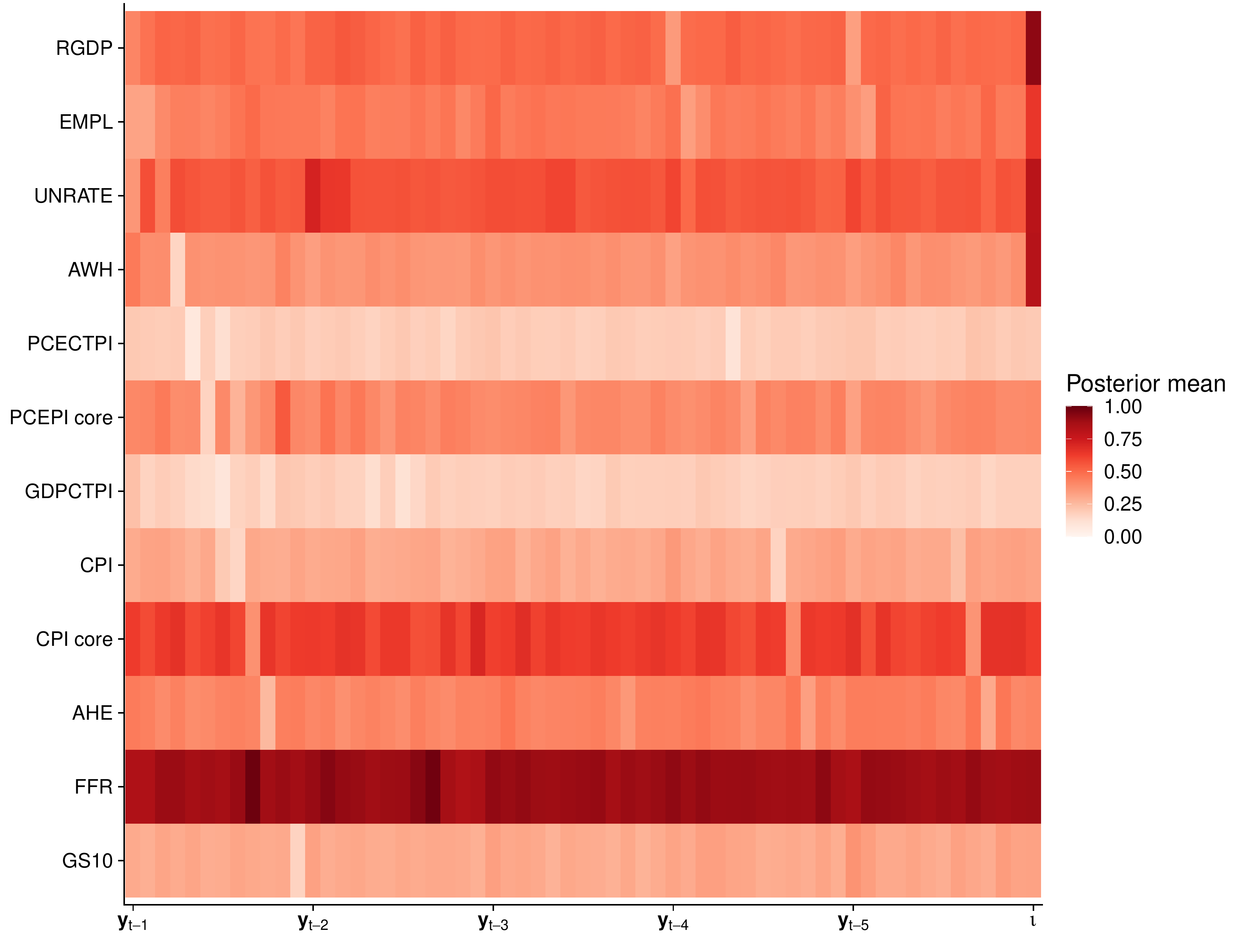}
\end{minipage}
\begin{minipage}{0.49\textwidth}
\centering
\includegraphics[width=1\textwidth,keepaspectratio]{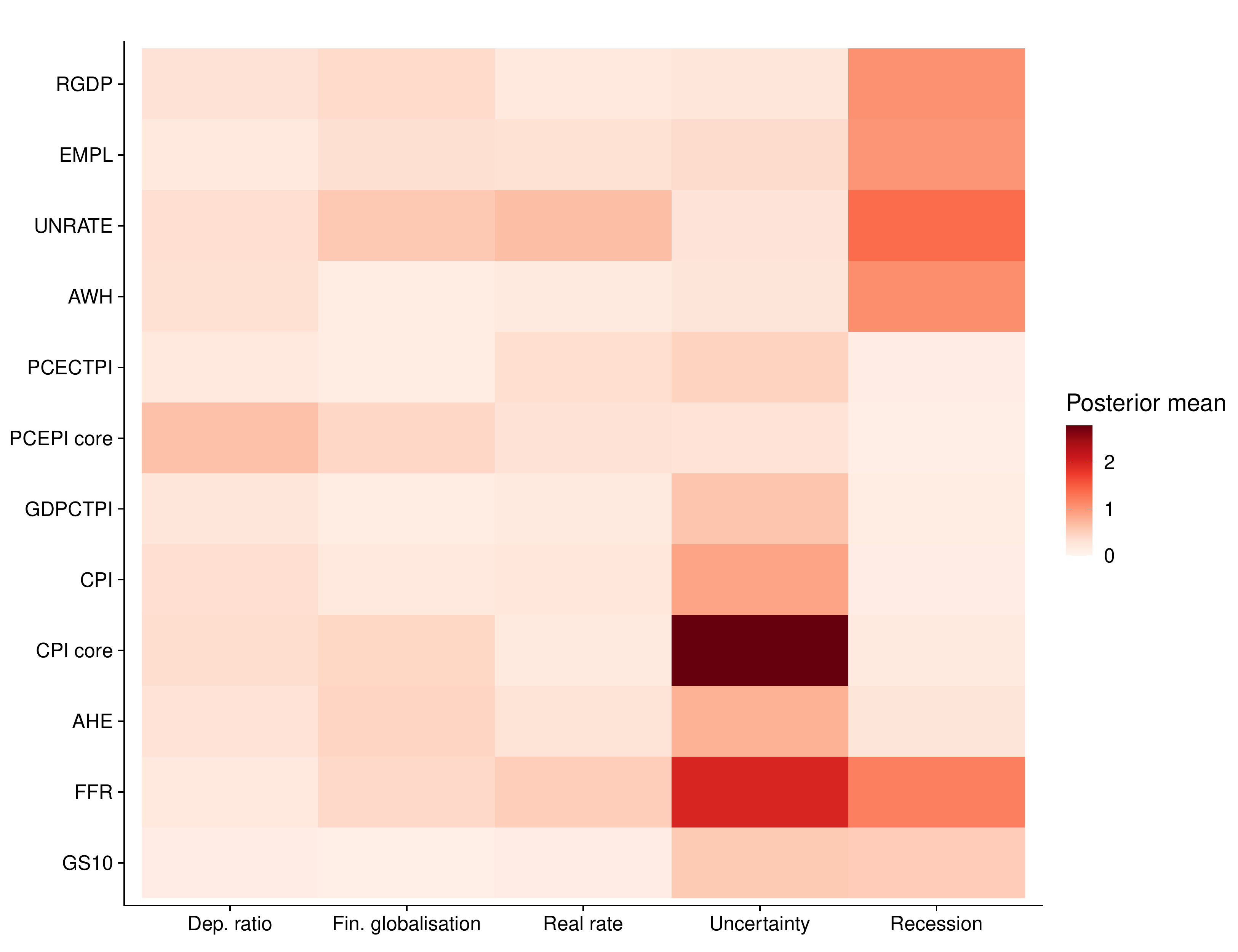}
\end{minipage}
\begin{minipage}{\textwidth}
\scriptsize \textbf{Notes:} Main specification with $Q_\beta = 25$ $(S_\beta = 1)$ for the conditional mean and $Q_q = 3$ $(S_q = 250)$ for the conditional variance-covariances. Panel (a) shows the posterior mean of the total share of time-variation explained for each coefficient. Vertical axis: endogenous variables. Front axis: coefficients related to the lags of $\bm y_t$. Panel (b) shows the posterior mean of the number of tree splits triggered by a certain effect modifier. For each effect modifier the number of splits is summed over the $Q_\beta (= 25)$ factors for the conditional mean. Vertical axis: endogenous variables. Front axis: effect modifiers.
\end{minipage}
\end{figure}

In this sub-section we consider what is driving the time variation in the VAR coefficients in our TVP-BART model with FHB.  \autoref{fig:insmpMEAN}(a) shows a heatmap of the total share of time-variation of $\bm \beta_{mt}$ explained by the nonlinear factors $F_m(\bm z_t)$ across equations $m=1, \dots, M$. This quantity, closely related to the familiar $R^2$, is computed as follows:
\begin{equation*}
 \text{diag}\left(\frac{\sum_{q=1}^{Q_\beta} \bm \lambda_{mq} \text{Var}(g(\bm z_t|\mathcal{T}^\beta_{mq}, \bm \mu_{mq}) \bm \lambda'_{mq}}{\sum_{q=1}^{Q_\beta} \bm \lambda_{mq} \text{Var}(g(\bm z_t|\mathcal{T}^\beta_{mq}, \bm \mu_{mq}) \bm \lambda'_{mq} + \bm V_m}\right),
\end{equation*}
with $\text{Var}(g(\bm z_t|\mathcal{T}^\beta_{mq}, \bm \mu_{mq})$ denoting the empirical variance of the function $g$. Dark red values indicate that a given TVP is driven almost exclusively by $F_m(\bm z_t)$, whereas white values suggest that most of the variation is driven by idiosyncratic movements in the TVPs.

Panel (b) of \autoref{fig:insmpMEAN} displays a heatmap of posterior means of the number of tree splits induced by one of the effect modifiers in $\bm z_t$ across coefficients and equations. This serves as a way to assess the relative importance of different effect modifiers in shaping the coefficient dynamics over time.

Starting with panel (a) of the figure, we see that the explanatory power of the TVP factors varies substantially across equations (and also across variables). While we find that TVPs in the interest rate and CPI core equations are strongly shaped by the effect modifiers, this share is considerably lower for the other equations. With two exceptions (PCETCPI and GDPCTPI), the shares are, however, sizable and often above 50 percent. Turning to PCETCPI and GDPCTPI, the effect modifiers explain a rather small amount of variation. Interestingly, for labor market quantities (EMPL, UNRATE, AWH) and real GDP we also find that the intercept (which determines the unconditional mean of the model) is strongly influenced by different effect modifiers. This indicates that long-run properties of these time series depend on covariates that may be interpreted as capturing structural change in the macroeconomy.

Focusing on panel (b) of the figure provides additional insights. First,  the old-age dependency ratio, financial globalization, and the real rate play only a limited role in explaining parameter dynamics. Second, for several variables we find that uncertainty shapes TVP dynamics. Among these are coefficients in the CPI and CPI core equations, the short-term interest rate equation, and the ten-year government bond yield. Third, for other variables such as output, employment, and the unemployment rate, we observe that uncertainty plays a more limited role. However, in these equations we instead find that the NBER's recession indicator is frequently included in the splitting rules.

\subsection{Capturing business cycle shocks}
 One of the main advantages of our nonparametric model is that, conditional on knowing the TVPs and error covariances, the model is a standard linear TVP-VAR model. Hence structural analysis, using identified impulse responses, can be readily carried out. In principle, an economist's preferred identification strategy based on, for example, sign restrictions \citep{benati2008}, zero impact restrictions \citep{primiceri2005, koop2009evolution}, or long-run restrictions can be implemented within our TVP-BART framework. 
 
 In this application, we focus on the question of how adverse business cycle shocks impact a set of inflation measures. To do so, we exploit the factor structure on the reduced-form VAR shocks to identify a business cycle shock \citep[for related identification approaches, see][]{korobilis2022new, chan2022large}. As emphasized by \cite{Gorodnichenko2005}, in VAR models like ours where the number of variables is relatively large (we have $M=12$) it can facilitate structural interpretation to have fewer structural shocks than $M$. In the next step, we trace out the dynamic evolution of our inflation measures to such a business cycle shock.
 
 One can decompose, as in \autoref{eq: factordecomposition}, the reduced-form VAR shocks into a factor component and an idiosyncratic measurement-error component (both of which are independent) under standard conditions \citep[see, for example,][]{anderson1956statistical, fruhwirth2018sparse, kaufmann2019bayesian}.\footnote{These conditions relate to the number of factors being smaller then the Ledermann bound and the number of non-zero elements in $\bm \Gamma$ being sufficiently large so that the decomposition in \autoref{eq: factordecomposition} is unique.} Absent heteroskedasticity, the resulting factors still have no economic interpretation and thus additional  structure is required to identify the shocks, given that the factors and the factor loadings can be rotated by any random orthogonal matrix. But, given the heteroskedasticity in $\bm R_t$, we can follow \cite{chan2022large} and identify, up to sign and scale, a business cycle shock as that factor (shock) that explains the largest amount of variation in innovations to output and unemployment variations during recessionary periods (as identified by the NBER). This identification strategy resembles the one proposed in \cite{bianchi2023inflation}. They identify business cycle shocks by searching for linear combinations of the reduced-form shocks of a trend-cycle VAR so as to maximize the amount of variation in unemployment or cyclical output.\footnote{Alternative approaches to identify business cycle shocks are proposed in \cite{del2020s} and \cite{Angeletos}.} 
 

 Specifically, our business cycle shock is obtained by computing:
 \begin{equation}
     \bm \zeta_{jt} = \text{diag}\left(\frac{r_{jt} \bm \gamma_j \bm \gamma'_j}{\bm \Gamma \bm R_t \bm \Gamma' + \bm \Sigma_t}\right), \label{eq:commonalities}
 \end{equation}
for all $j$ and finding that factor that maximizes the variances explained for real GDP and the unemployment rate during recessionary episodes.  This yields, for each MCMC draw, a factor that can be interpreted as a business cycle shock. To point-identify the sign of the factors and the associated loadings, we normalize the factors and loadings to identify the business cycle shock as having a negative impact effect on output growth and a positive impact effect on unemployment. 

In summary, we identify the business cycle shock and the associated impulse responses via the following steps:
\begin{enumerate}[leftmargin=*]
    \item We identify the business cycle factor by finding the factor that explains the largest amount of variation in output growth and unemployment during NBER-defined recessions; see \autoref{eq:commonalities}. Without loss of generality, let us assume that this factor is the $j^{th}$ element of $\bm q_t, q_{jt}$, and the corresponding loadings are $\bm \gamma_j$.
    \item We compute the responses to a unit increase in $q_{jt}$. The impact reaction of $y_t$ is given by the loadings $\bm \gamma_j$ and higher order impulse responses are computed using standard recursions based on the companion form of the VAR at time $t$.
    \item To identify whether the shock is contractionary or expansionary (that is, the sign of $\bm \gamma_j$), we check the impact reaction for real output growth and the unemployment rate. If the former decreases and the latter increases, we label the shock as being contractionary. 
    \item Finally, to anchor the magnitude of the shock, we normalize the impact responses such that real output growth declines by one standard deviation on average.
\end{enumerate}
These steps yield partial identification, implying that the business cycle shock is uniquely identified whereas the remaining $(Q_q - 1)$ factors (and the associated columns in $\bm \Gamma$) are left unrestricted. This identification approach is related to ones developed in recent papers \citep{korobilis2022new, chan2022large} which advocate using sign restrictions on the factor loadings to pin down a shock of interest. But our approach differs in the sense that we solve the column switching problem (which is required to attach an economic meaning to the different factors) through a narrative approach that builds on the notion that business cycle shocks are the ones that determine the largest amount of variation in real activity quantities during recessions. Our approach could easily be combined with sign-restricted factor stochastic volatility models, by introducing certain restrictions on the prior associated with $\bm \Gamma$.

\begin{figure}[!htbp]
\centering
\caption{\textbf{Shock labeling.} Share of the variance of business cycle variables explained by factors over time. \label{fig:insmpVAR}}
\begin{minipage}{\textwidth}
\end{minipage}
\begin{minipage}{\textwidth}
\centering
\includegraphics[width=0.8\textwidth,keepaspectratio]{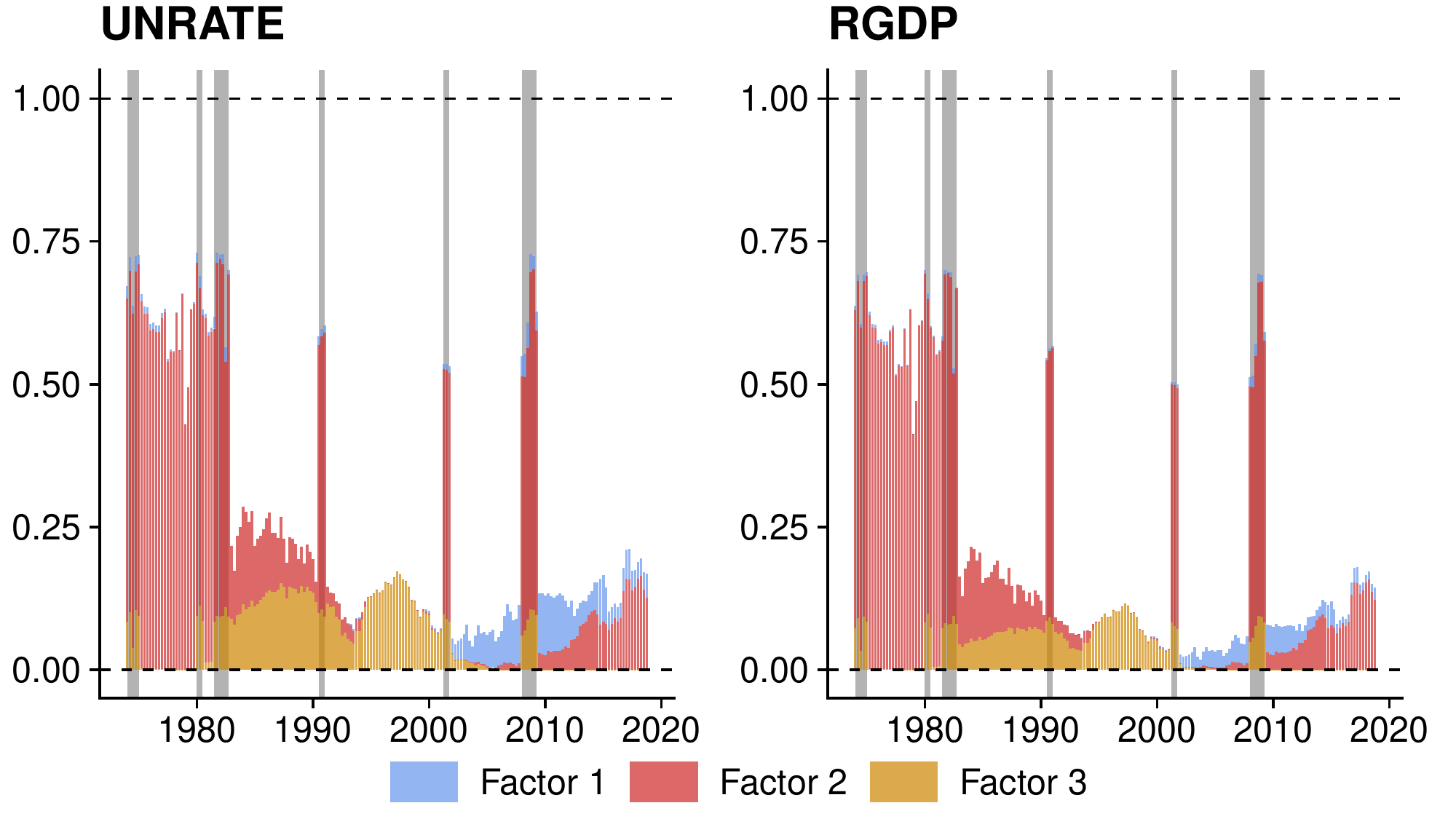}
\end{minipage}
\begin{minipage}{\textwidth}
\scriptsize \textbf{Notes:} Main specification with $Q_\beta = 25$ $(S_\beta = 1)$ for the conditional mean and $Q_q = 3$ $(S_q = 250)$ for the conditional variance-covariances. This figure shows the posterior mean of the share of the variance of business cycle variables explained by factors over time (by focusing on the diagonal elements of $\bm \Omega_t$ related to unemployment and real output growth). We then refer to the factor that accounts for the maximal volatility of real output growth and unemployment as our business cycle shock. Vertical axis: share explained by each factor. Front axis: quarters.
\end{minipage}
\end{figure}

\autoref{fig:insmpVAR} plots the posterior mean of the proportion of the variation, $\bm \zeta_{jt}$, in the unemployment rate and in output growth explained by the three factors over time. This figure shows that the second factor explains the largest amount of variation in the early part of the sample (until the twin recession of the early 1980s) and during all recessions in our sample. During recessions, this factor explains close to 70 percent of the variation in the reduced-form shocks to both unemployment and output growth. 

\subsection{Impulse responses to a business cycle shock}
In this sub-section, we look at the dynamic effects of our business cycle shock. Impulse response functions are computed by shocking the business cycle (second) factor and tracing out the dynamic reactions of $y_{t+h}$ for $h=1, \dots, 16$. 

\begin{figure}[!htbp]
\centering
\caption{Average impulse response over horizons.\label{fig:IRFhetero}}
\includegraphics[width=1\textwidth,keepaspectratio]{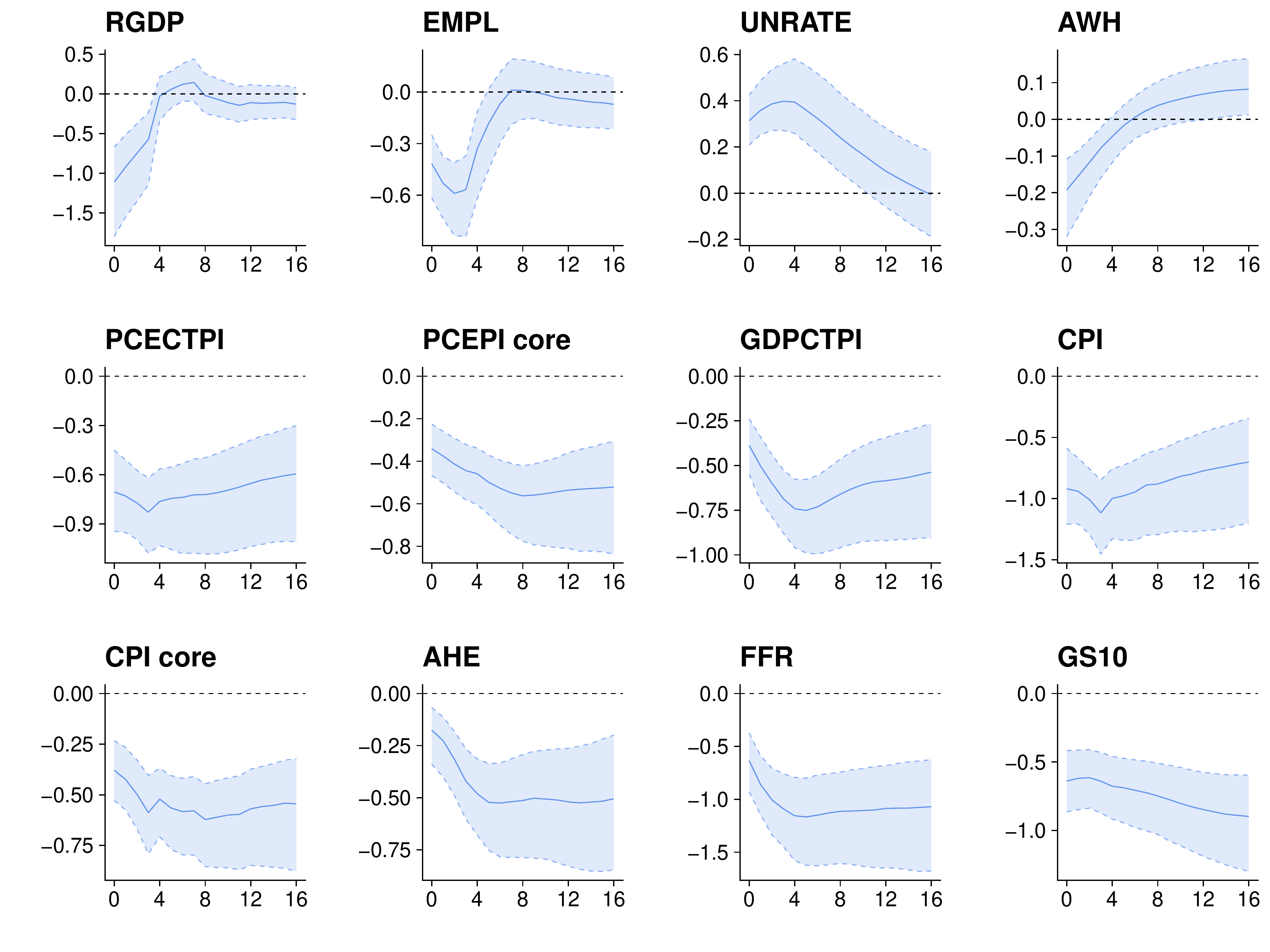}
\begin{minipage}{\textwidth}
\scriptsize \textbf{Notes:}  Impulse responses to a negative business cycle shock, averaged over time. Blue solid lines denote the posterior median, blue dashed lines the $16^{th}$/$84^{th}$ posterior percentiles, with the blue shaded areas corresponding to the $68\%$ credible sets, and the black dashed lines mark the zero line. Panels: endogenous variables. Vertical axis: impulse responses. Front axis: horizons (in quarters).
\end{minipage}
\end{figure}

Since our model features TVPs, the impulse responses can be computed at each point in time. This gives us a posterior distribution over $T$ period-specific IRFs, a statistical object that is difficult to visualize. To aid exposition, we start our analysis by considering average impulse responses. These are obtained by averaging the time-specific impulse responses over time and are depicted in \autoref{fig:IRFhetero}.

\autoref{fig:IRFhetero} shows that, averaged over time, a contractionary business cycle shock leads to unemployment rising and inflation (including core and wage inflation) falling. The dynamic effects on the different inflation measures are similar, but long-lasting. Like the main business shock of  \cite{Angeletos}, the peak effect of our business cycle shock on the real variables  also occurs within a year or two. Specifically, we observe that output and employment decline while the unemployment rate increases. Real GDP growth reacts rapidly by declining by around one percentage point on impact. For employment growth, the peak effect materializes after about three quarters. The unemployment rate quickly increases and displays a peak reaction of around 0.5 percentage points after around one year. These reactions are largely consistent (both in terms of shape and size) with the ones reported in \cite{bianchi2023inflation}.

\begin{figure}[!htbp]
\centering
\caption{Average Phillips curve multipliers over horizons.\label{fig:PChetero}}
\includegraphics[width=1\textwidth,keepaspectratio]{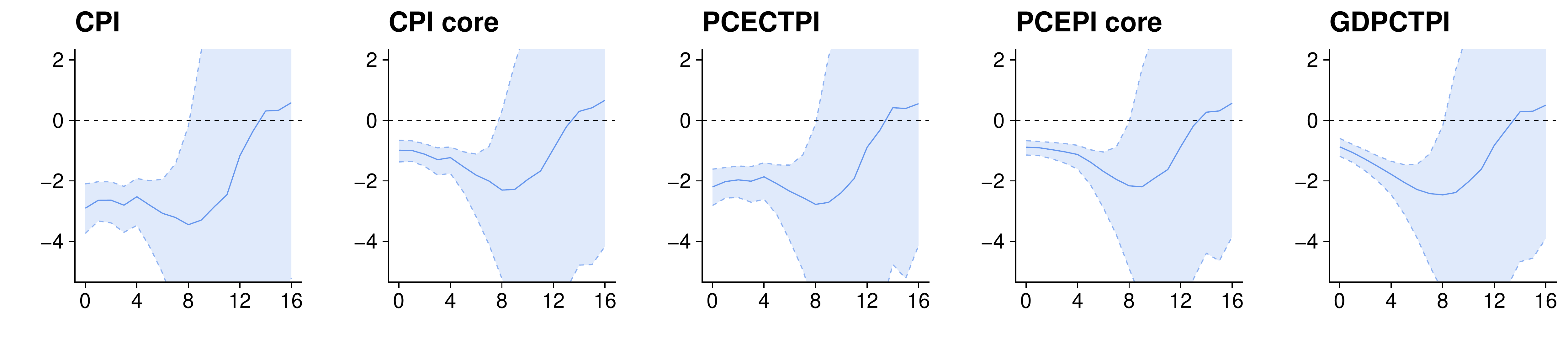}
\begin{minipage}{\textwidth}
\scriptsize \textbf{Notes:}  Philips curve multipliers based on a negative business cycle shock, averaged over time. Blue solid lines denote the posterior median, blue dashed lines the $16^{th}$/$84^{th}$ posterior percentiles, with the blue shaded areas corresponding to the $68\%$ credible sets, and the black dashed lines mark the zero line. Panels: price indices. Vertical axis: impulse responses. Front axis: horizons (in quarters).
\end{minipage}
\end{figure}

\begin{figure}[!htbp]
\centering
\caption{Impulse responses over time at the one-year-ahead horizon ($h = 4$). \label{fig:IRFthetero}}
\begin{minipage}{0.49\textwidth}
\centering
\includegraphics[width=0.7\textwidth,keepaspectratio]{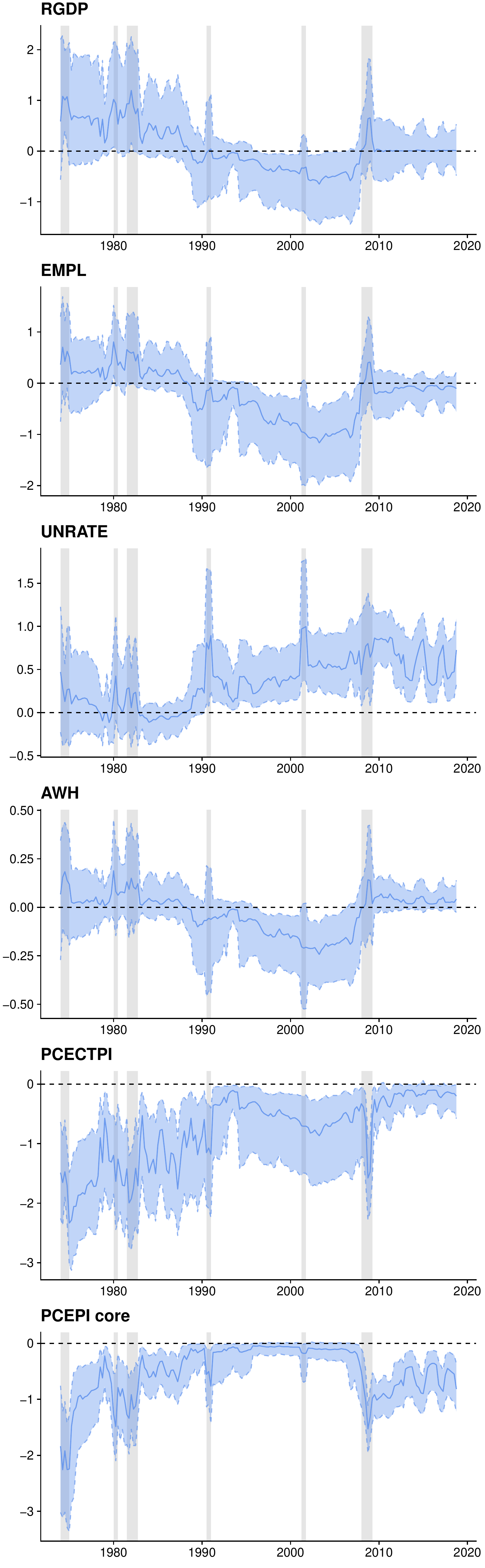}
\end{minipage}
\begin{minipage}{0.49\textwidth}
\centering
\includegraphics[width=0.7\textwidth,keepaspectratio]{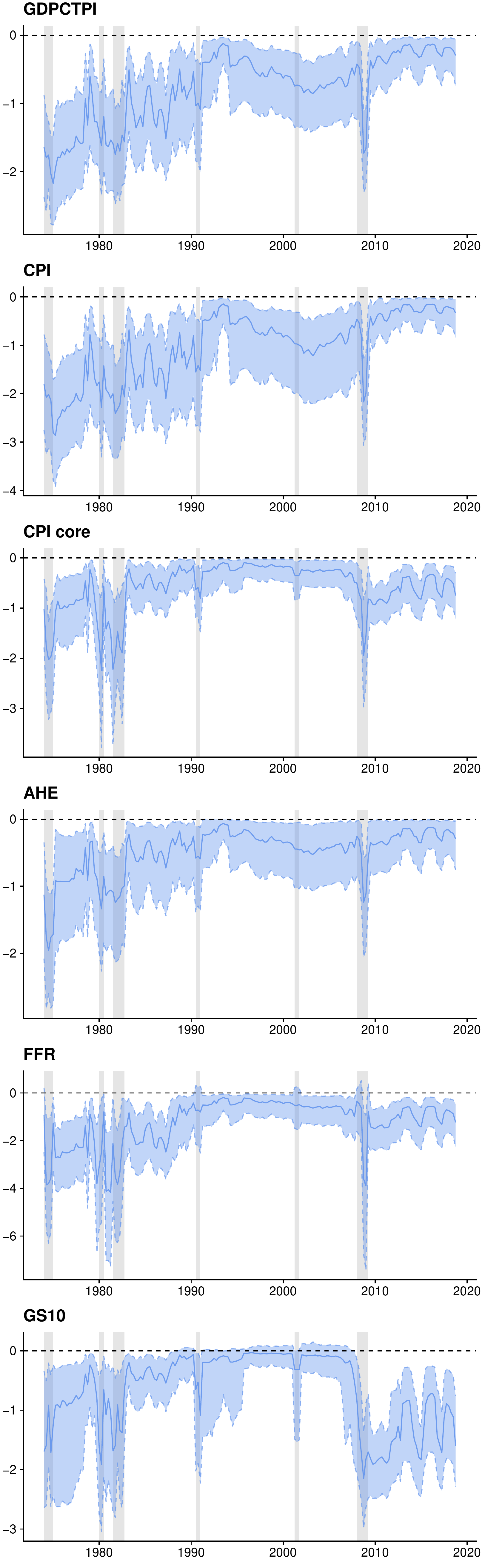}
\end{minipage}
\begin{minipage}{\textwidth}
\scriptsize \textbf{Notes:} Time-specific impulse responses to a negative business cycle shock. Blue solid lines denote the posterior median, blue dashed lines the $16^{th}$/$84^{th}$ posterior percentiles, with the blue shaded areas corresponding to the $68\%$ credible sets, and the black dashed lines mark the zero line. Panels: endogenous variables. Vertical axis: impulse responses. Front axis: periods (in quarters).
\end{minipage}
\end{figure}

When we consider the reactions of our different inflation measures, we find that prices decline on impact. This reduction appears to be quite persistent. As we will show below (see \autoref{fig:IRFthetero}), this persistent reaction of prices is mainly driven by strong and persistent declines of inflation up to the early 1990s. These results are consistent with the existence of a negatively sloped Phillips curve --- at least on average through the $1975$:Q$1$ to $2019$:Q$4$ period.

To hone in on this relationship between inflation and unemployment, we normalize the IRFs of the different price measures by the IRFs of the unemployment rate. \cite{barnichon2021phillips} call this quantity the Phillips curve multiplier. The multipliers are shown in \autoref{fig:PChetero}. We again see that, on average over time, the Phillips curve multipliers are negative and statistically significant. \autoref{fig:PChetero} also reveals that these negative effects persist for two to three years. And they vary by inflation measure. As we should expect, the Phillips curve is stronger for headline than core measures of inflation. The strongest effects on inflation are typically seen two years after the business cycle shock.

To understand to what degree averaging over time is masking temporal variations in the Phillips curve relationship, \autoref{fig:IRFthetero} plots, at the one-year-ahead horizon ($h=4$), the impulse responses due to the contractionary business cycle shock at each point in time. The ability to identify and capture structural change of different forms is a key feature of our model. \autoref{fig:IRFthetero} reveals that there are indeed important temporal variations. The responses of, in particular, the headline inflation measures become more muted over time. Focusing in on the effects on CPI inflation, we see that the business cycle shock lowers inflation significantly through the $1970$s and $1980$s. But the responses thereafter are more muted. They become increasingly muted as we look to the period after the global financial crisis. Interestingly, evidencing a clear nonlinearity, there is a strong negative effect on inflation during the recessionary period associated with the global financial crisis itself. Our findings therefore provide ex-post justification for the decision by \cite{del2020s} to estimate their VAR model, designed to understand the Phillips curve, on samples before and after $1990$. But our results also reveal important temporal instabilities and changes within these two periods that are lost by simple sample-slit or indeed rolling regressions as also often used in the literature.

Turning to the effects on unemployment, again consistent with \cite{del2020s}, \autoref{fig:IRFthetero} shows that the response of unemployment to a business cycle shock becomes more persistent over time. This is consistent with economic expansions lasting longer in more recent decades. But \autoref{fig:IRFthetero} adds texture to this narrative by revealing that recessionary periods, except for $2008$-$9$, are marked by especially strong responses.

Bringing together the price inflation and unemployment responses, we conclude that the sensitivity of price inflation to unemployment has weakened markedly since $1990$. The response of wage inflation to the the business cycle shock is weaker throughout the sample. This casts doubt on the view \citep[see, for example,][]{Knotek, HOOPER2020} that the Phillips curve is stronger for wage than for price inflation. Since the $1990$s the impulse responses for wage inflation and price (CPI) inflation look broadly similar; see  \autoref{fig:IRFthetero}. This includes evidence that wage as well as price inflation did decline in response to a business cycle shock during the Great Recession, with prices declining by more than wages.

\subsection{Scenario analysis to assess the channels of time variation}\label{sec:scenario}
\begin{figure}[!htbp]
\centering
\caption{Impulse response of prices for different scenarios at the one-year-ahead horizon ($h = 4$). \label{fig:IRFscenhetero}}
\includegraphics[width=0.9\textwidth,keepaspectratio]{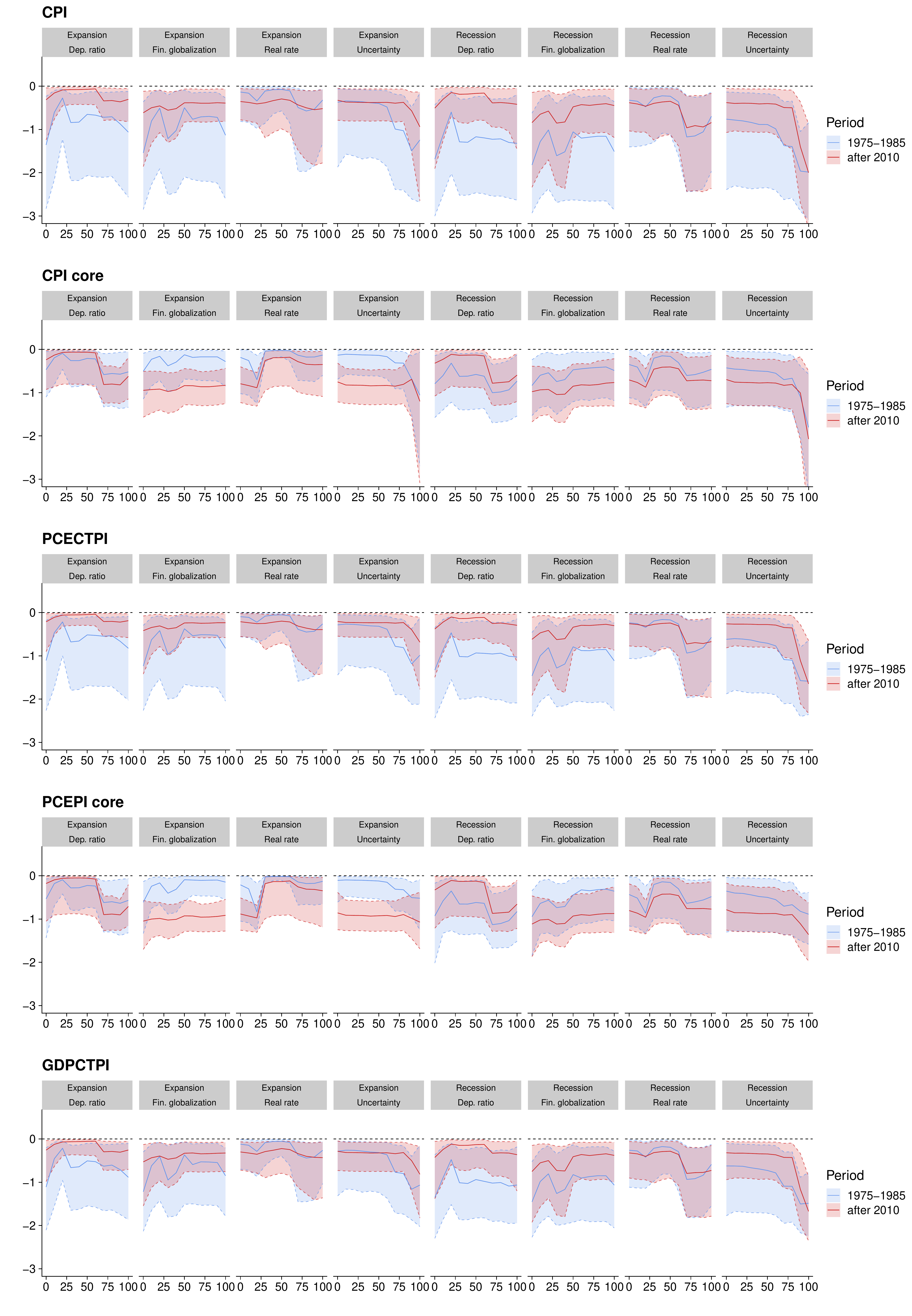}
\begin{minipage}{\textwidth}
\scriptsize \textbf{Notes:} Impulse responses to a negative business cycle shock by partially varying the effect modifiers. For example, the top-left panel refers to the responses across percentiles of the dependency ratio while assuming an expansion state and setting the remaining effect modifiers (that is, financial globalization, real interest rate, and uncertainty) either to the mean of the subsample of periods $1975$:Q$1$ to $1984$:Q$4$ (colored in blue) or to the mean of the subsample of periods from $2010$:Q$1$ to $2019$:Q$4$ (colored in red).
Colored solid lines denote the posterior median, colored dashed lines the $16^{th}$/$84^{th}$ posterior percentiles, with the colored shaded areas corresponding to the $68\%$ credible sets, and the black dashed lines mark the zero line. Vertical panels: price indices. Horizontal panels: effect modifiers for an expansion and a recession state. Vertical axis: impulse responses. Front axis: percentiles in $\%$ where $0\%$ $(100\%)$ denotes the minimum (maximum) value.
\end{minipage}
\end{figure}

One key feature of our model is that it allows us to link the time variation in the parameters (and thus functions thereof such as IRFs) to the effect modifiers in $\bm z_t$. Since $\bm z_t$ influences the TVPs using a nonparametric model, it is difficult to clearly answer how changes in $\bm z_t$ impact the TVPs. Since our interest centers on the implied IRFs, we can, however, carry out simulations that show how the dynamic responses to a business cycle shock change as we vary $\bm z_t$. The results of this exercise are shown in \autoref{fig:IRFscenhetero}. This figure depicts the price responses in the rows of the panel and in the columns shows different assumptions on $\bm z_t$. 

To analyze whether IRFs differ in expansion and recessions, we set the NBER recession indicator to zero  (that is, we assume that the economy is in an expansion) or to one (that is, we assume that the economy is in a recession). Based on this, we vary one of the effect modifiers while setting the remaining effect modifiers to some pre-specified value. This pre-specified value is either the average value over the period $1975$ to $1985$ (which are the blue-shaded IRFs in the figure) or the period after $2010$ (which are the red-shaded IRFs). This allows us to capture the general macroeconomic environment in the respective time periods. This gives us four overall combinations for the IRFs. We consider how prices react in expansions and recessions and whether there are discernible differences in the transmission of business cycle shocks in these two regimes. Based on one of these four general scenarios, we set each effect modifier (for example, the dependency ratio, financial globalization, the real rate, and uncertainty) equal to different sample quantiles and then compute the implied price IRFs. This provides a detailed picture on how impulse responses depend on the effect modifiers.

\autoref{fig:IRFscenhetero} confirms that the headline (non-core) inflation measures were more strongly affected by business cycle shocks before 1985. The most striking nonlinearity for the effect modifiers is seen with respect to uncertainty. As uncertainty increases beyond its $75^{th}$ percentile, we see much stronger negative effects on all the inflation measures, including with post-2010 data. This effect is especially pronounced during recessionary periods. This all supports a view that the Phillips curve remains alive and well during times of recession and greater-than-average uncertainty, events that empirically tend to co-exist. This is consistent with theories of the financial accelerator, suggesting that shocks have amplified effects in recessions.

\section{Conclusion}\label{sec:conclusions}
In this paper, we have developed a nonparametric model that uses Bayesian additive regression trees (BART) methods to allow for change of an unknown form in both the conditional means and variances of a multivariate time series model (a VAR). Unlike existing nonparametric approaches, interpretation and macroeconomic inference including structural analysis is easier, since, as the model gives a nonparametric treatment to the parameters rather than the variables, it remains conditionally linear in the mean. An additional novel feature is that the new model allows for nonparametric factor structures for parameters in the conditional means and variances, thus reducing the number of nonparametric functions to estimate and ensuring parsimony. 

In an empirical exercise we show how the proposed nonparametric VAR model contributes to our understanding of the time-varying nature of the Phillips curve. Inflation has become considerably less sensitive to business cycle shocks, in particular since $1990$. However, the flexible nonlinear features of the model show that the effects on inflation remain strong when uncertainty rises to high levels.

\footnotesize{\setstretch{0.9}
\addcontentsline{toc}{section}{References}
\bibliography{tvpbart}
\bibliographystyle{cit_econometrica.bst}
}\normalsize

\clearpage

\newpage \normalsize
\setcounter{page}{1}
\setcounter{footnote}{0}
\begin{appendices}
\begin{center}
\LARGE \textbf{Online Appendices}  \\
 \Large{\textbf{Bayesian Modeling of TVP-VARs \\[-0.5em] 
 Using Regression Trees}} \\
\vspace*{10pt}
\normalsize \uppercase{Niko Hauzenberger,$^{1,2}$} \uppercase{Florian Huber,$^1$} \\[-0.5em] 
\uppercase{Gary Koop},$^2$ and \uppercase{James Mitchell}$^3$\\
\vspace*{10pt}
\textit{$^1$University of Salzburg}\\[-0.5em]
\textit{$^2$University of Strathclyde}\\[-0.5em]
\textit{$^3$Federal Reserve Bank of Cleveland}\\
\vspace*{20pt}
\date{}
\end{center}

\setcounter{equation}{0}
\setcounter{table}{0}
\setcounter{figure}{0}
\renewcommand\theequation{A.\arabic{equation}}
\renewcommand\thetable{A.\arabic{table}}
\renewcommand\thefigure{A.\arabic{figure}}
\renewcommand\thesubsection{A.\arabic{subsection}}
\section{Additional details about the prior}\label{app:A}
In the main body of the paper we have focused on the prior for the tree part of the model. The remaining priors are relatively standard in the literature. 

For the time-invariant VAR coefficients ($\bm A$), the factor loadings associated with the static error factors ($\bm \Gamma$), and the factor loadings in the state equation ($\bm \Lambda_m$) we use a Horseshoe prior. Let $\phi_{ij}$ denote a generic coefficient. A general Horseshoe prior is given by:
\begin{equation*}
    \phi_{ij}|\varrho_{ij} \varpi \sim \mathcal{N}(0, \varrho_{ij}^2 \varpi^2), \quad \varrho_{ij} \sim \mathcal{C}^+(0, 1), \quad \varpi \sim \mathcal{C}^+(0, 1),
\end{equation*}
where $\varrho_{ij}$ is a coefficient-specific shrinkage parameter that forces each coefficient to zero and $\varpi$ denotes an equation-specific shrinkage parameter that forces all elements in the respect coefficient matrix ($\bm A, \bm \Lambda_m$, or $\bm \Gamma$) to zero. This prior differs across parameter types in one important way. For the VAR coefficients, the global shrinkage parameter applies to all elements in $\bm A$. For the factor loadings, we estimate separate global shrinkage parameters for each column of $\bm \Lambda_m$ and $\bm \Gamma$, respectively. This enables us to determine the number of factors by forcing columns to zero, thus excluding a specific factor from the model.

\clearpage
\setcounter{equation}{0}
\setcounter{table}{0}
\setcounter{figure}{0}
\renewcommand\theequation{B.\arabic{equation}}
\renewcommand\thetable{B.\arabic{table}}
\renewcommand\thefigure{B.\arabic{figure}}
\renewcommand\thesubsection{B.\arabic{subsection}}
\section{Data Appendix}\label{app:C}
\input{data.tex}

\setcounter{equation}{0}
\setcounter{table}{0}
\setcounter{figure}{0}
\renewcommand\theequation{C.\arabic{equation}}
\renewcommand\thetable{C.\arabic{table}}
\renewcommand\thefigure{C.\arabic{figure}}
\renewcommand\thesubsection{C.\arabic{subsection}}
\section{Empirical Appendix}\label{app:B}
\subsection{How BART can approximate TVPs: An Illustration}\label{app:illust}
In this sub-section, we provide a simple illustration of how BART works when used for approximating TVPs. We use a single equation example, with $y_t$ being quarterly US CPI inflation and $x_t$ being the unemployment rate.\footnote{Details about the data used in this empirical illustration are provided in Section 5. The model is estimated using the MCMC methods described in Section 3.2 of the main paper.}  Hence, $M=K=1$. Moreover, we set $Q_\beta=Q_q=S_q=1$ and focus this illustration on the role the number of trees in the conditional mean  ($S_\beta$) plays in modeling the TVPs. Thus, there is a single TVP which is the Phillips curve coefficient (that is, the coefficient on the unemployment rate). We compare our estimate of this using various choices for $S_\beta$ to the estimates produced by a standard TVP regression, where the Phillips curve coefficient evolves according to a random walk. To aid in comparability of the Philips curve coefficient across models, every model includes an intercept that evolves according to a random walk. 

Under these assumptions our model  reduces to:
\begin{equation*}
\begin{aligned}
y_t &= c_t + \beta_t x_t + (q_t + \varepsilon_t), \quad (q_t + \varepsilon_t) \sim \mathcal{N}(0, R(z_t) + \sigma^2_t),\\
\beta_t &= F(z_t) + \eta_t, \quad \eta_t \sim \mathcal{N}(0, v^2),
\end{aligned}    
\end{equation*}
where $c_t$  has a random walk state equation. The conventional TVP regression replaces the second equation with a random walk but otherwise is identical. 

We illustrate our BART-based techniques with $z_t = t$, so that the only explanatory variable in BART is a deterministic trend. In this case, the splitting rules would divide the time periods into distinct regimes that would feature their own terminal node parameters. The resulting model can be interpreted as a regime switching model with an unknown number of regimes and a diagonal transition probability matrix. This specification is closely related to the TVP regression model developed in \cite{hauzenberger2021fast}, which uses sparse finite mixtures to model the time-variation in the coefficients.

\begin{figure}[!htbp]
\centering
\caption{Estimated  branching structure for the Phillips curve coefficient using a single tree. \label{fig:trees}}
\includegraphics[width=1\textwidth,keepaspectratio]{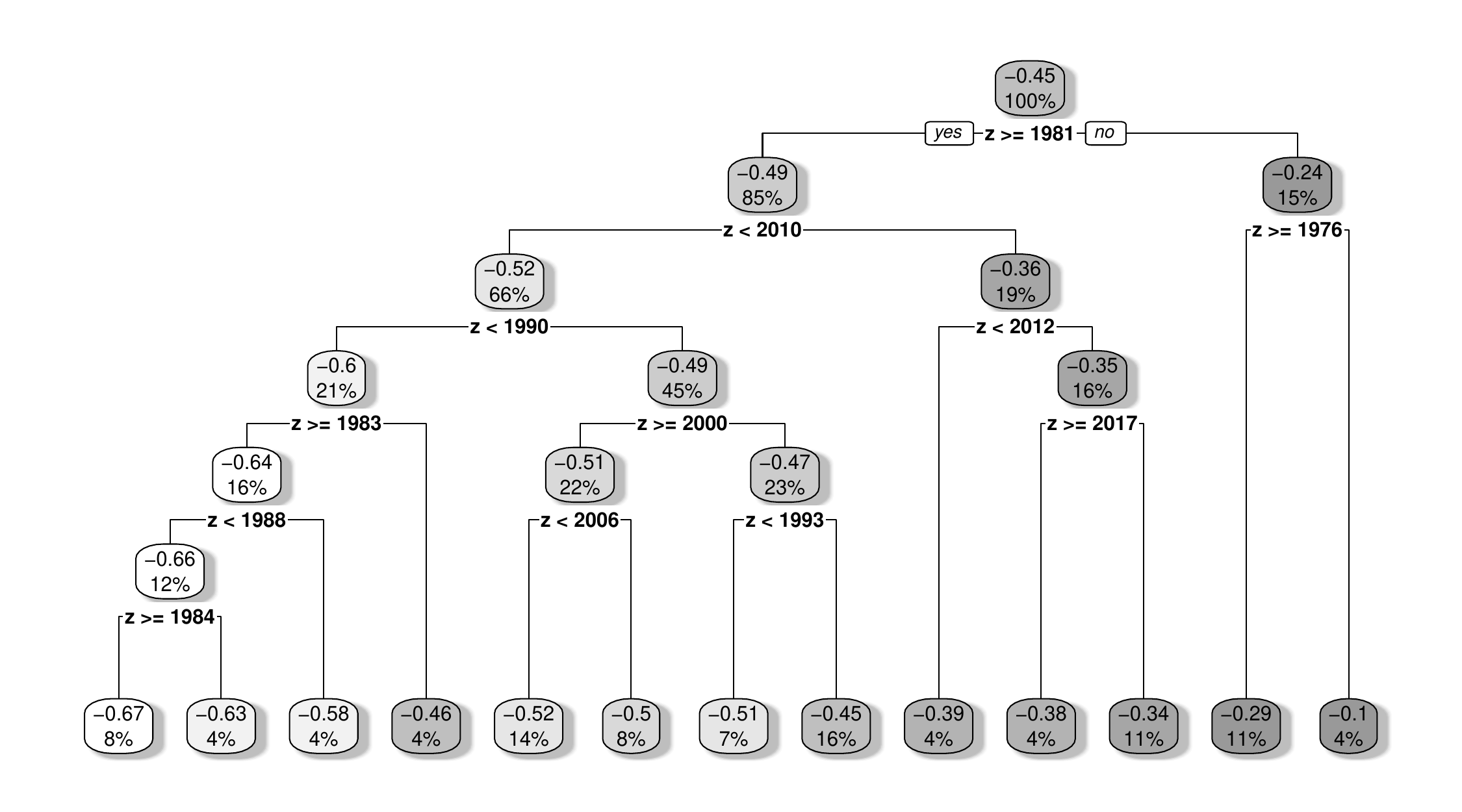}
\begin{minipage}{\textwidth}
\scriptsize \textbf{Notes:} As the exogenous effect modifier ($\bm z$), we use a simple linear time trend ranging from $1974$:Q$1$ to $2019$:Q$4$. Each oval box indicates the terminal node parameter of a particular branch and the share (in percent) of observations belonging to this branch. The splitting rules effectively decompose the time-varying coefficient into distinct regimes, featuring their own terminal node parameters. 
\end{minipage}
\end{figure}

Figure \ref{fig:trees} shows the single estimated tree we obtain when we set $S_\beta=1$. The oval boxes in the bottom row of the figure contain estimated values of the Phillips curve coefficient produced by the tree (along with the percentage of observations which share each value). Since we have set $z_t=t$ the tree divides up observations into different time periods. Hence, we basically have a regime structure where different regimes have different Phillips curve coefficients. For instance, eight percent of the observations have a Phillips curve coefficient which is very negative ($-0.67$). Proceeding from the top of the tree, it can be seen that these observations occur in the interval $[1984,1988)$.

\begin{figure}[!htbp]
\centering
\caption{Estimates of the time-varying Phillips curve coefficient using a single tree, five trees ($S_\beta= 5$), and many trees ($S_\beta= 150$).\label{fig:beta}}
\includegraphics[width=1\textwidth,keepaspectratio]{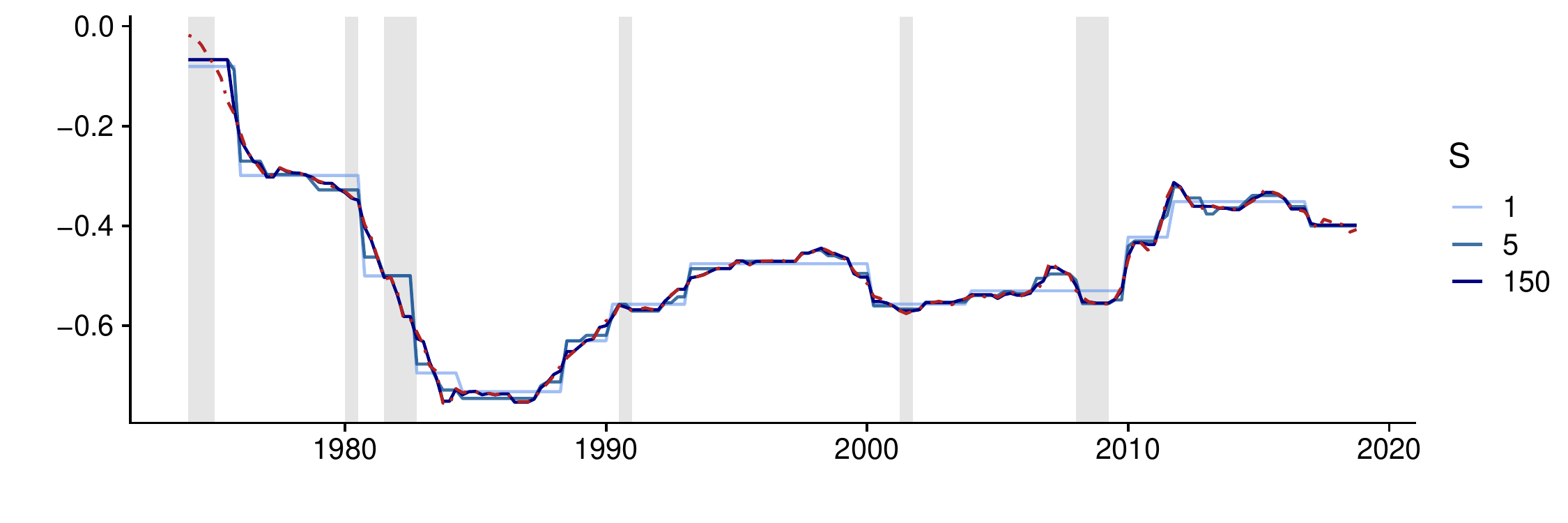}
\begin{minipage}{\textwidth}
\scriptsize \textbf{Notes:} The blue lines indicate estimates of $\beta_t$ for a varying number of trees, while the red dash-dotted line refers to estimates of $\beta_t$ assuming random walk evolution of $\beta_t$. These models are estimated as special cases of our general model using the MCMC algorithm outlined in Sub-section \ref{sec:MCMC}.
\end{minipage}
\end{figure}

The regime structure can also be seen in \autoref{fig:beta}, which plots different estimates of $\beta_t$, including the case above with  $S_\beta=1$. In particular, it plots the (smoothly evolving) estimate produced by the TVP regression with random walk evolution of the coefficients against estimates produced by a single tree, five trees ($S_\beta=5$), and many trees ($S_\beta=150$). All four lines in the figures are similar to one another, but it can be seen that adding more trees allows for more regimes and the fitted line produced by BART becomes smoother and tracks the random walk evolution of the parametric model more and more closely.

\subsection{Assessing model fit using information criteria}\label{app:model_fit}
To assess model adequacy and decide on some parameters of the model, we focus on the widely applicable/Watanabe-Akaike information \citep[WAIC,][]{watanabe2010asymptotic, watanabe2013widely}. The WAIC is a generally applicable measure of model fit that, on the one hand, rewards model fit while, on the other hand, punishing model complexity. We opt for the WAIC due to its excellent characteristics when applied to tightly parameterized hierarchical models.\footnote{\cite{gelman2014understanding} and \cite{vehtari2017practical} thoroughly study the characteristics of the WAIC.} Following \cite{gelman2014understanding}, the WAIC for model $\textit{A}$ is given by:
\begin{equation*}
\text{WAIC}_A = -2 \left(\widehat{\text{lpd}}_A - \hat{p}_A \right). 
\end{equation*}
Here, $\widehat{\text{lpd}}_A $ denotes an estimate of the log point-wise predictive density: 
\begin{equation*}
   \widehat{\text{lpd}}_A  = \sum_{t = 1}^{T} \log \left(\frac{1}{S} \sum_{s = 1}^{S} p_A(\bm y_t|\bm \Theta_A^{(s)})  \right),
\end{equation*}
where $\bm \Theta_A^{(s)}$ is generic notation that refers to the $s^{th}$ draw of the parameters and  latent states from the full posterior distribution of model $\textit{A}$. This term measures model fit. Model complexity is measured through  $\hat{p}_A$. This quantity  can be interpreted as the effective number of parameters and is given by the variance of the point-wise log likelihood across $S$ draws from the full posterior distribution: 
\begin{equation*}
   \hat{p}_A  = \sum_{t = 1}^{T} \text{Var} \left(\log p_A(\bm y_t|\{\bm \Theta_A^{(s)}\}_{s = 1}^{S}) \right).
\end{equation*}

We compare our nonparametric TVP-FBART-FHB to alternatives that retain factor structures for the TVPs and multivariate stochastic volatility processes but are parametric.\footnote{We do not compare our approach to TVP-VARs lacking such factor structures, such as the model of \cite{primiceri2005}, since they would be heavily over-parameterized in models of our dimension. Thus, all of the models in our comparison adopt factor structures to ensure parsimony.} Specifically, we compare our approach to the model proposed in \cite{chan2020reducing} using the factor stochastic model  proposed in \cite{kastner2020sparse}. This model has a similar structure to ours, involving a factor structure both in the TVPs and the multivariate stochastic volatility process. We assume random walk behavior for the factor driving both the TVPs and the stochastic volatility process. We use the abbreviation TVP-FRW-FSV for this model. We emphasize that our approach differs from TVP-FRW-FSV only in modeling time variation nonparametrically, instead of with random walks.  We also consider a constant-coefficient VAR with FSV. 

In Table \ref{tab:WAIC} we compute WAICs for TVP-FBART-FHB and TVP-FRW-FSV for different numbers of factors ($Q_\beta$ and $Q_q$) as well as, for the BART-based approaches, the number of trees driving the TVPs ($S_\beta$). This lets us pin down a preferred model specification and investigate the relationship between $Q_\beta$ and $S_\beta$. The trade-off between $Q_\beta$ and $S_\beta$ is potentially interesting. In principle, setting $Q_\beta$ to a large value and $S_\beta$ to a small value leads to a model which is closely related to a standard BART specification for TVP-VARs (that is, many factors driving the TVPs but each is relatively simple, involving a small number of trees). In contrast,  a model that sets $Q_\beta$ to a small value and $S_\beta$ to a large one implies the TVPs are driven by a small number of factors, but these factors are potentially very complicated involving a large number of trees. We investigate this relationship by considering different combinations of $Q_\beta$ and $S_\beta$ that reflect a range of cases. Since this relationship is less relevant in the lower dimensional error covariance matrix and we have found more robustness to choice of number of trees (provided it is not too small), we follow the literature on heteroBART and set $S_q=250$.

\input{WAIC.tex}
Table \ref{tab:WAIC} suggests that performance of the TVP-FBART-FSV model strongly depends on the choices of $Q_\beta$, $Q_q$, and $S_\beta$. The best performance is obtained by setting both the number of factors in the conditional mean to be large (but with a small number of trees) and the number of factors in the FHB process to be small. Setting the number of factors in the FHB part of the model to be large leads to a deterioration in performance, regardless of the choices made relating to the FHB model. This suggests a fair degree of common volatility, but less commonality for the conditional mean coefficients. 

We do see some trade-off between the number of factors and the number of trees in the FBART part of the model. Models with many factors and few trees perform best, but models with fewer factors and more trees perform nearly as well. The worst performance is found for models with a moderate number of factors and trees. 

The fact that the constant coefficient VAR with FSV can be beaten by TVP models if the number of factors is chosen judiciously indicates that there is parameter change in the VAR coefficients. But the fact that the TVP-FBART-FHB models nearly always exhibit substantially lower WAIC values than the TVP-FRW-FSV models with the same number of factors suggests that the parameter change is better modeled nonparametrically than via random walks.  

\clearpage

\subsection{Two-year-ahead impulse responses to a business cycle shock}\label{app:irfs}
\begin{figure}[!htbp]
\centering
\caption{Impulse responses over time at the two-year-ahead horizon ($h = 8$).}
\begin{minipage}{0.49\textwidth}
\centering
\includegraphics[width=0.7\textwidth,keepaspectratio]{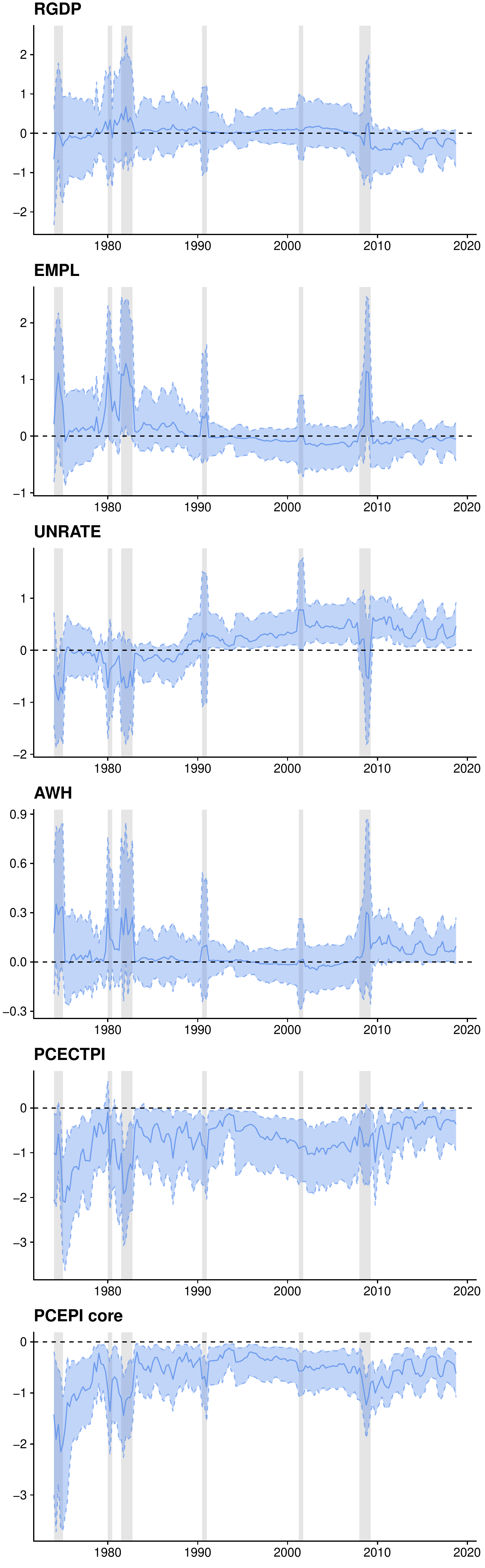}
\end{minipage}
\begin{minipage}{0.49\textwidth}
\centering
\includegraphics[width=0.7\textwidth,keepaspectratio]{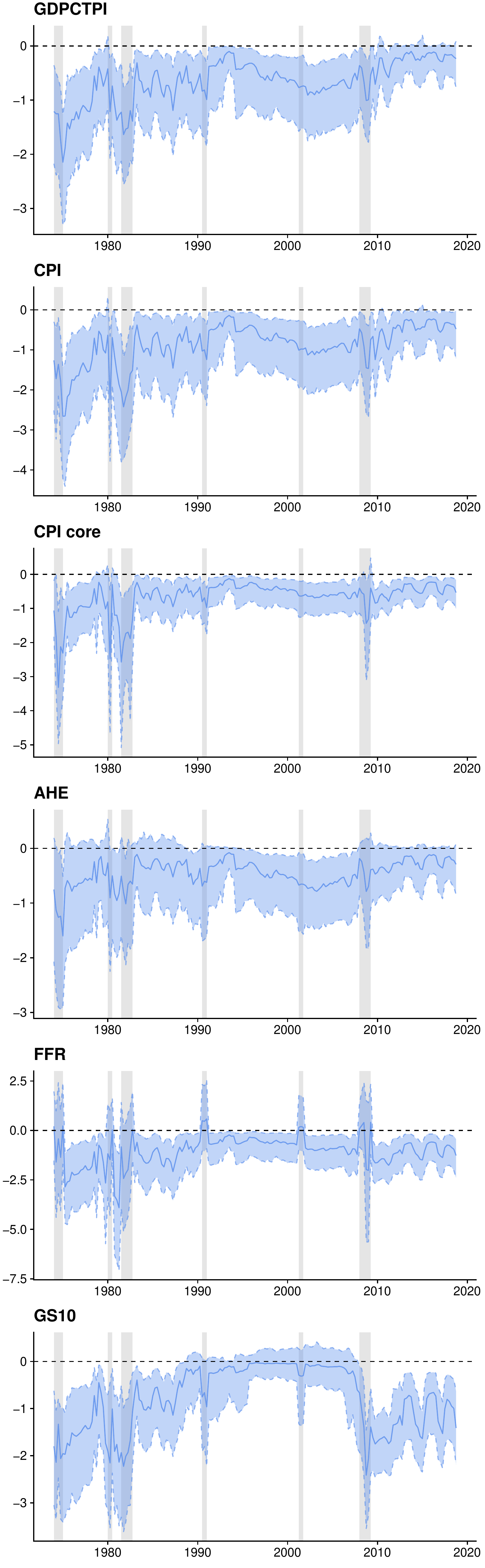}
\end{minipage}
\begin{minipage}{\textwidth}
\scriptsize \textbf{Notes:}  Time-specific impulse responses to a negative business cycle shock. Blue solid lines denote the posterior median, blue dashed lines the $16^{th}$/$84^{th}$ posterior percentiles, with the blue shaded areas corresponding to the $68\%$ credible sets, and the black dashed lines mark the zero line. Panels: endogenous variables. Vertical axis: impulse responses. Front axis: periods (in quarters).
\end{minipage}
\end{figure}

\begin{figure}[!htbp]
\centering
\caption{Impulse response of prices for different scenarios at the two-year-ahead horizon ($h = 8$).}
\includegraphics[width=0.9\textwidth,keepaspectratio]{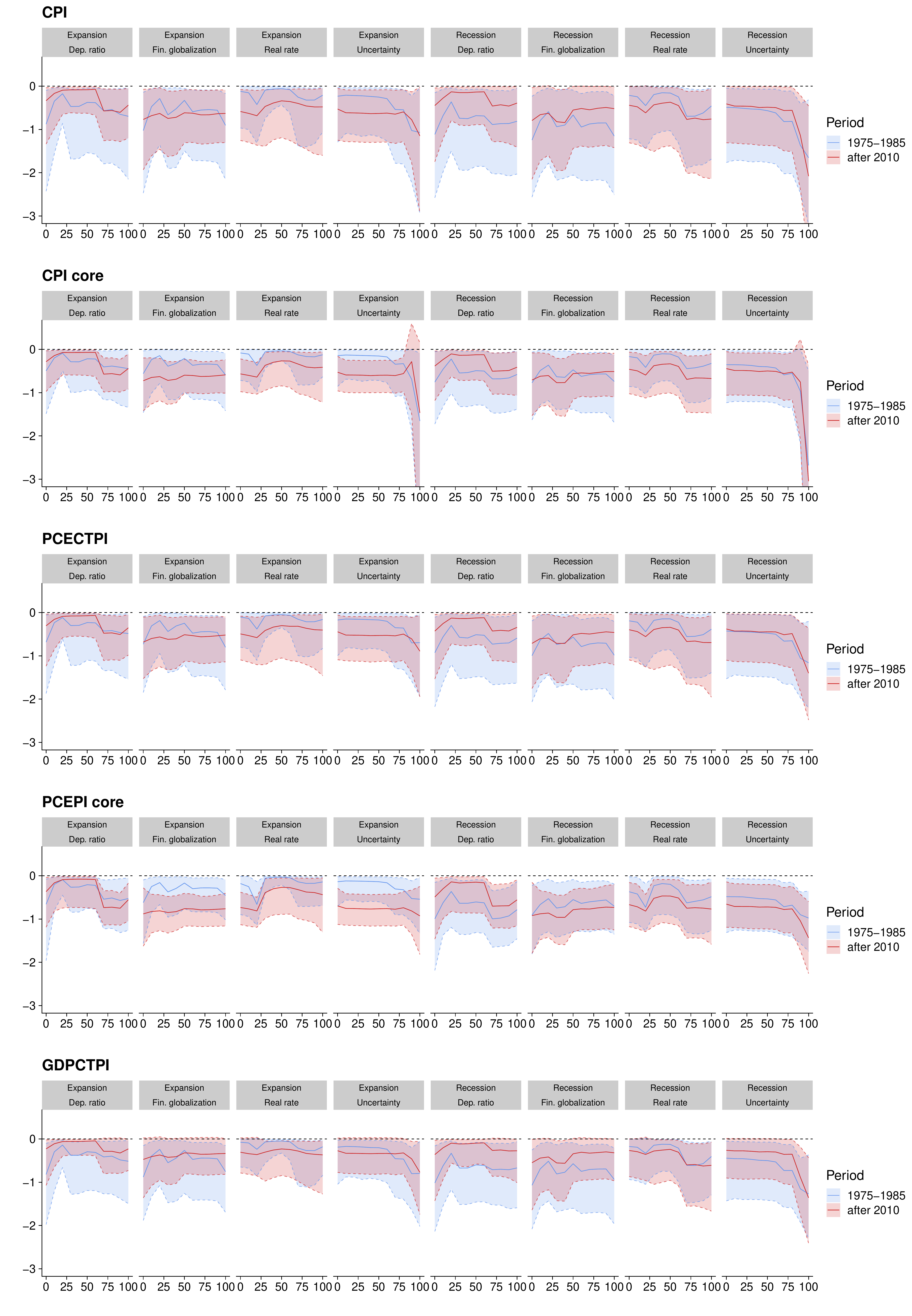}
\begin{minipage}{\textwidth}
\scriptsize \textbf{Notes:} Impulse responses to a negative business cycle shock by partially varying the effect modifiers. For example, the top-left panel refers to the responses across percentiles of the dependency ratio while assuming an expansion state and setting the remaining effect modifiers (i.e., financial globalization, real interest rate, and uncertainty) either to the mean of the subsample of periods $1975$:Q$1$ to $1984$:Q$4$ (colored in blue) or to the mean of the subsample of periods from $2010$:Q$1$ to $2019$:Q$4$ (colored in red).
Colored solid lines denote the posterior median, colored dashed lines the $16^{th}$/$84^{th}$ posterior percentiles, with the colored shaded areas corresponding to the $68\%$ credible sets, and the black dashed lines mark the zero line. Vertical panels: price indices. Horizontal panels: effect modifiers for an expansion and a recession state. Vertical axis: impulse responses. Front axis: percentiles in $\%$ where $0\% (100\%)$ denotes the minimum (maximum) value. 
\end{minipage}
\end{figure}

\end{appendices}
\end{document}

%% file: data.tex
\begin{table}[!htbp]
{\tiny
\caption{Data description. \label{tab:data}} 
\begin{tabular}{llllll}
\toprule
& \textbf{Label}  & \textbf{Mnemonic} & \textbf{Description} & \textbf{Transformation} &  \\ 
\midrule
\multicolumn{6}{c}{(a) Endogenous variables $\bm y_t$}\\[0.5em]
& RGDP &     GDPC1            &  Real gross domestic product          &          y-o-y growth rate      &  \\
& EMPL &     CE16OV           &  Employment        &              y-o-y growth rate   &  \\
& UNRATE &     UNRATE           &  Unemployment rate        &         none       &  \\
& AWH &     CES0600000007    &  Avg. weekly hours (production and nonsuperv. employees) & none &  \\
& PCECTPI     &      PCECTPI         &  Chain-type price index: personal consumption expenditures         &           y-o-y growth rate      &  \\
& PCEPI core  &    PCEPILFE          &  Personal consumption expenditures excl. food and energy       &          y-o-y growth rate       &  \\
& GDPCTPI      &     GDPCTPI          &  Chain-type price index: gross domestic product          &             y-o-y growth rate    &  \\
& CPI          &       CPIAUCSL       &  Consumer price index (all items)        &                y-o-y growth rate  &  \\
& CPI core     &   CPILFESL           &  Core consumer price index (all items less food and energy)          &           y-o-y growth rate      &  \\
& AHE          &      CES0600000008   & Avg. hourly earnings (production and nonsuperv. employees)  & y-o-y growth rate  &  \\
&  FFR          &      FEDFUNDS        & Effective federal funds rate &  none  &  \\
& GS10         &     GS10             & $10$-year government bond yields    & none   &  \\
\midrule
\multicolumn{6}{c}{(b) Exogenous effect modifiers $\bm z_t$}\\[0.5em]
& Dep. ratio       & -- & Age dependency ratio: older dependents to working-age population&  none \\
& Fin. globalization & -- & KOF financial globalization index & none  \\
& Real rate   & -- &  \cite{laubach2003measuring} estimate of the natural rate of interest &  none  \\
& Recession & -- & National Bureau of Economic Research (NBER) recession indicator  & none \\
& Uncertainty & -- & \cite{jurado2015measuring} estimate of macroeconomic uncertainty & none  \\
\bottomrule
\end{tabular}}
\begin{minipage}{\textwidth}
\vspace{0.1cm}
\scriptsize \textbf{Notes:} For the endogenous variables we rely on the \cite{mccracken2016fred} data set, while for the exogenous effect modifiers we draw on different data sources. The age dependency ratio and the NBER recession indicator are obtained from the FRED database of the Federal Reserve Bank of St. Louis (\href{https://fred.stlouisfed.org}{fred.stlouisfed.org}), the financial globalization indicator is downloaded from KOF Swiss Economic Institute at the ETH Zurich (\href{https://kof.ethz.ch/en/forecasts-and-indicators/indicators/kof-globalisation-index.html}{kof.ethz.ch/en/forecasts-and-indicators/indicators/kof-globalisation-index.html}), the real rate estimate of \cite{laubach2003measuring} from the database of the Federal Reserve Bank of New York (\href{https://www.newyorkfed.org/research/policy/rstar}{www.newyorkfed.org/research/policy/rstar}), and the macroeconomic uncertainty from the personal webpage of Sydney C. Ludvigson (\href{https://www.sydneyludvigson.com/macro-and-financial-uncertainty-indexes}{sydneyludvigson.com/macro-and-financial-uncertainty-indexes}). Since both the age dependency ratio and the financial globalization indicator are available only on an annual basis, they are interpolated to the quarterly frequency using a Kalman smoother.
\end{minipage}
\end{table}

%% file: WAIC.tex
\begin{table}[!htbp]
\footnotesize{
\caption{Model comparison based on relative WAICs. \label{tab:WAIC}} 
\begin{tabular*}{\textwidth}{l @{\extracolsep{\fill}} cccccccc}
\toprule
\multicolumn{1}{l}{\bfseries VAR specification}&\multicolumn{1}{c}{}&\multicolumn{2}{c}{\bfseries Conditional mean}&\multicolumn{1}{c}{}&\multicolumn{4}{c}{\bfseries Conditional variances}\tabularnewline
\multicolumn{5}{c}{}&\multicolumn{4}{c}{Number of factors $Q_q$}\tabularnewline
\cmidrule{3-4} \cmidrule{6-9}
\multicolumn{2}{c}{}&\multicolumn{1}{c}{$Q_\beta$}&\multicolumn{1}{c}{$S_\beta$}&\multicolumn{1}{c}{}&\multicolumn{1}{c}{1}&\multicolumn{1}{c}{3}&\multicolumn{1}{c}{6}&\multicolumn{1}{c}{12}\tabularnewline
\midrule
~~TVP-BART-FHB&&1&100&&0.80&0.99&1.52&1.61\tabularnewline
~~&&2&50&&0.85&0.95&1.26&1.61\tabularnewline
~~&&5&20&&1.02&1.00&1.21&1.52\tabularnewline
~~&&10&10&&1.49&1.11&1.41&1.47\tabularnewline
~~&&10&1&& 0.69&0.78&1.22&1.50\tabularnewline
~~&&25&1&&0.82& \cellcl 0.69&1.10&1.55\tabularnewline
\midrule
~~TVP-FRW-FSV&&1&&&1.10&1.41&1.78&1.76\tabularnewline
~~&&2&&&0.94&1.32&1.51&1.54\tabularnewline
~~&&5&&&0.81&1.22&1.37&1.40\tabularnewline
~~&&10&&&0.74&1.16&1.21&1.21\tabularnewline
\midrule
~~constant with FSV&&&&&0.95&1.09&1.65& \cellclbench 1.00\tabularnewline
\bottomrule
\end{tabular*}
\begin{minipage}{\textwidth}
\vspace{0.1cm}
\scriptsize \textbf{Notes:} TVP-FBART-FHB refers to the nonparametric model with a factor BART form for the conditional mean and factor heteroBART for the conditional variance. TVP-FRW-FSV denotes the parametric model with TVP and SV processes following factor random walks. The red shaded cell indicates the best model specification, while the gray shaded cell indicates the benchmark (a constant-coefficient VAR with FSV and $Q_q = 12$). The WAIC is evaluated jointly for unemployment (UNRATE), overall consumer price inflation (CPIAUCSL), and the federal funds rate (FEDFUNDS). 
\end{minipage}}
\end{table}